\documentclass[12pt]{iopart}
\usepackage{iopams}
\usepackage{setstack}
\usepackage{amssymb}

\usepackage{graphicx}
\usepackage[dvips]{color}
\usepackage{dcolumn}
\usepackage{bm}

\def\<{\langle}
\def\>{\rangle}

\newcommand{\arrowspace}{\hspace{-0.1cm}}
\newcommand{\hH}{\hat{H}}
\newcommand{\hV}{\hat{V}}
\newcommand{\hc}{\hat{c}}
\newcommand{\hb}{\hat{b}}
\newcommand{\ha}{\hat{a}}
\newcommand{\hd}{\hat{d}}
\newcommand{\hK}{\hat{K}}
\newcommand{\hphi}{\hat{\phi}}
\newcommand{\heta}{\hat{\eta}}
\newcommand{\hpsi}{\hat{\psi}}
\newcommand{\hrho}{\hat{\rho}}
\newcommand{\hPsi}{\hat{\Psi}}
\newcommand{\teins}{$|\mathrm{t}_1\rangle_\sigma$}
\newcommand{\tzwei}{$|\mathrm{t}_2\rangle_\sigma$}
\newcommand{\tdrei}{$|\mathrm{t}_3\rangle_\sigma$}
\newcommand{\eeins}{$|\mathrm{e}_1\rangle$}
\newcommand{\ezwei}{$|\mathrm{e}_2\rangle_{\sigma\sigma}$}
\newcommand{\geins}{$|\mathrm{g}_1\rangle$}
\newcommand{\gzwei}{$|\mathrm{g}_2\rangle_{\sigma\sigma}$}
\newcommand{\gdrei}{$|\mathrm{g}_3\rangle$}
\newcommand{\sgn}{\mathrm{sgn}}
\newcommand{\gate}{\mathrm{gate}}
\newcommand{\bias}{\mathrm{bias}}
\newcommand{\leads}{\mbox{\small leads}}
\newcommand{\edge}{e}
\newcommand{\bulk}{b}
\newcommand{\Sct}[1]{Sec.\ \ref{#1}}

\newcommand{\Eq}[1]{Eq.\ (\ref{#1})}
\newcommand{\Eqs}[2]{Eqs.\ (\ref{#1})\ and\ (\ref{#2})}
\newcommand{\Tab}[1]{Tab.\ \ref{#1}}
\newcommand{\Tabs}[2]{Tabs.\ \ref{#1}, \ref{#2}}
\newcommand{\Fig}[1]{Fig.\ \ref{#1}}

\newcommand{\App}[1]{\ref{#1}}
\newcommand{\eb}{\mathrm{e-b}}
\newcommand{\bb}{\mathrm{b-b}}
\begin{document}
\title{Spin-dependent transport through interacting graphene armchair nanoribbons}
\author{Sonja Koller$^{1}$, Leonhard Mayrhofer$^{1,2}$, and Milena Grifoni$^{1}$}

\address{$^1$ Theoretische Physik, Universit\"at Regensburg, 93040 Germany\\
$^{2}$ Fraunhofer IWM, W\"ohlerstra\ss e 11, 79108 Freiburg, Germany}


%
\date{\today}
%
\begin{abstract}
We investigate spin effects in transport across fully interacting, finite size graphene armchair nanoribbons (ACNs) contacted to collinearly spin-polarized leads.
In such systems, the presence of short-ranged Coulomb interaction between bulk states and states localized at the ribbon ends leads  to novel spin-dependent phenomena.
Specifically, the total spin of the low energy many-body states is conserved during tunneling but that of the bulk and end states is not.  As a consequence, in the single-electron regime, dominated by Coulomb blockade phenomena, we find pronounced negative differential conductance features for ACNs contacted to parallel polarized leads. These features are however absent  for an anti-parallel contact configuration, which in turn leads within a certain gate and bias voltage region to a negative
tunneling magneto-resistance.
 Moreover, we analyze the changes in the transport characteristics under the influence of an external
magnetic field.
\end{abstract} 
%
\maketitle

\section{Introduction}

\begin{figure}
\begin{center}
\includegraphics[width=8.8cm]{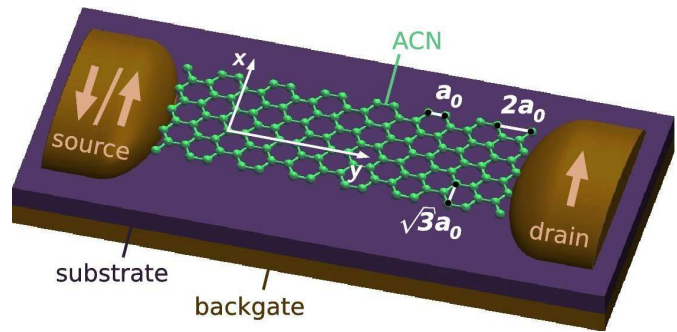}
\caption{\label{graphene}A graphene armchair nanostripe contacted by ferromagnetic leads. At the long sides, the lattice is terminated in armchair, at the small ends in zig-zag configuration. The length of a bond between two carbon atoms is $a_0\approx 0.14\,$nm. We choose the orientation of the coordinate system such that the $x$-axis points along the zig-zag ends, the $y$-axis along the long armchair edges of the stripe.}
\end{center}
\end{figure}

Since the first successful separation of a one atom thick sheet of graphite by Novoselov and Geim and coworkers~\cite{Novoselov04}, graphene, i.e. an isolated single sheet of graphite, has been a material attracting ever raising interest.  Not only a great potential for applications~\cite{Falkol07,Chen08}, but also fundamental physics issues~\cite{BeenakkerCOL} arise from the linear dispersion relation in the electronic band structure of graphene, predicted about sixty years ago~\cite{Wallace47},
 as confirmed by various recent experimental findings~\cite{Novoselov04,Zhang05,Zhou06,Bostwick07}.\\
Increasing effort is presently put in the understanding of the
electronic properties of graphene nanodevices, which can
be obtained by etching or lithographic techniques and may achieve lateral dimensions of a few tenth of nm~\cite{Han07,Stampfer08,Lin08}.
Studies on the effects of electron-electron interactions and confinement in transport across graphene nanodevices have been carried out on single-dots~\cite{Stampfer08,Schnez09} and, only recently, on double-dot structures~\cite{Molitor09,Moriyama09}. A desirable goal is the fabrication of clearly defined geometries, and of particular interest for applications~\cite{Falkol07} could be narrow stripes of graphene, so-called carbon nanoribbons. Conductance quantization has been observed in 30nm wide ribbons~\cite{Lin08}, while an energy gap near the charge neutrality point scaling with the inverse ribbon width was reported in~\cite{Han07} and theoretically~\cite{Sols07,Zarea07} attributed to Coulomb interaction effects.\\

 Crucial for the properties of a nanoribbon is the form of its ends. The two most regular possibilities are an armchair and a zig-zag end (see Fig. \ref{graphene}). At present, the shape of the ends cannot be controlled, but there is ongoing progress in developing methods to fabricate stripes with clear edges, by scanning tunneling microscope litho\-graphy~\cite{Tapaszto08}, chemical synthesis~\cite{Yang08} or by unrolling carbon nanotubes~\cite{Jioa09}. Moreover, there exist theoretical studies~\cite{Nakada96,Akhmerov08} claiming that any narrow stripe should show either the behavior of a zig-zag or of an armchair ribbon (ACN), where the names specify the form of the long side edges.\\A peculiar property of the zig-zag edge is the existence of localized states~\cite{Fujita96} which were indeed observed experimentally by means of scanning tunneling microscopy~\cite{Klusek05,Kobayashi06,Niimi06}. Due to their high degeneracy, flat-band ferro\-magnetism is expected from the Hubbard model, leading to a spin-polarized many particle ground state. It was suggested~\cite{Son06,Wimmer08} to exploit this property for spintronic applications, where transport is governed by carriers in the oppositely polarized channels along the two long zig-zag edges.
 Recently both Hubbard and long-ranged Coulomb interaction effects  have been analyzed \cite{Wunsch08} under the assumption of a filled valence and an empty conduction
band (half-filling).
There was a prediction of strong spin features in case of a low population of the localized mid-gap states.\\
In contrast, in armchair ribbons the localized states are at the far apart zig-zag ends. As we showed recently \cite{Koller09}, for narrow armchair ribbons short-ranged Coulomb interaction, i.e. local scattering events between either the two sublattices of graphene or between the extended bulk and the localized end states, gain increasing weight. They lead to an exchange coupling, inducing entanglement between the bulk and the end states. This has a decisive impact on eigenstates and transport properties.\\

In this work we extend the investigations of Ref. \cite{Koller09} to the case in which the ACN is contacted to ferromagnetic leads and/or additionally subjected to a magnetic field. Various remarkable features due to the presence of the entangled end-bulk states, e.g. a negative tunneling magneto-resistance in a fully symmetric set-up, are discussed.\\
The paper is structured as follows: In \Sct{sec:Noninteracting-electrons-in} and \Sct{sec:interaction-ham}, we present our low energy theory for \emph{narrow} ACNs.
Readers not interested in technical details can, on the basis of the results for the ACN Hamiltonian summarized in \Sct{sec:diagdiag}, directly start from \Sct{sec:minmod}, where we shortly analyze a minimal set of states relevant for the explanation of features in transport across ACNs. Specifically, we show that scattering between end and bulk electrons causes an entanglement of states with the same total spin-projection $S_z$, but different configurations of end and bulk spins. As a consequence, neither for the spins trapped at the ribbon ends, nor for the bulk electrons, $S_z$ is a good quantum number any more, which could be a handicap for proposed quantum information applications~\cite{Falkol07}. Although the formation of states symmetric or antisymmetric under the exchange of end spins remains without a special signature in the spectrum, it in fact leaves strong fingerprints in transport, as discussed in Ref. \cite{Koller09} for an unpolarized set-up as well as in \Sct{sec:transport} for collinear contact magnetizations. The most prominent feature in spin-dependent transport for collinear lead magnetizations is a negative tunneling magneto-resistance (TMR) within a narrow region along the edges of the Coulomb diamonds for even fillings, \Sct{sec:pol}. Finally we explain how the transport characteristic is expected to change under application of an external magnetic field, both for non-magnetic and collinearly polarized contacts, \Sct{sec:mag}.

\section{Noninteracting electrons in a finite size graphene armchair nanoribbon\label{sec:Noninteracting-electrons-in}}
\begin{figure}
\begin{center}
\includegraphics[width=5.8cm]{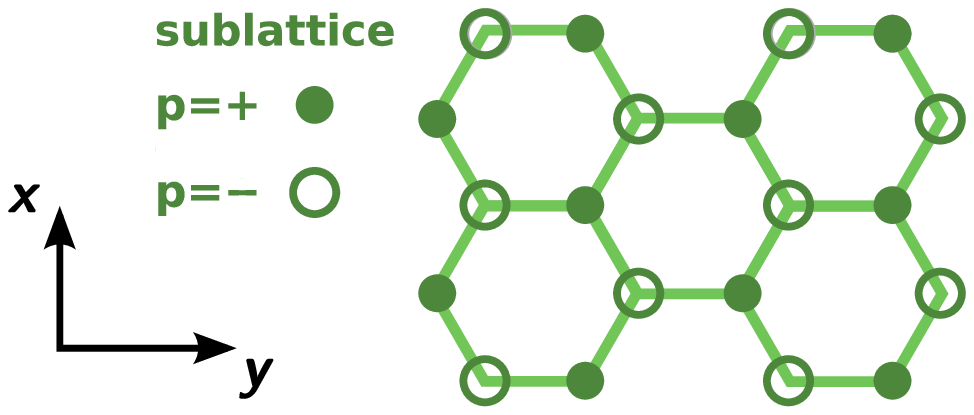}
\caption{\label{lattice}The sublattice structure of the graphene honeycomb lattice. The underlying Bravais lattice is triangular, with a basis of two sites labelled $p=\pm$.}
\end{center}
\end{figure}
Being crucial for the subsequent analysis, we review here the end and bulk states of \emph{narrow}, noninteracting ACNs.\\
The carbon atoms in the graphene lattice are arranged in a hexagonal honeycomb
lattice. There are two atoms per unit cell such that we can define
two different sublattices $p=\pm$ as shown in Fig. \ref{lattice}. Hybridization
of the $2s$-orbital with the $2p_{x}$- and $2p_{y}$-orbitals leads
to strong $\sigma$-bonds in the lattice plane. The electrons in the
remaining $2p_{z}$ orbitals form $\pi$-bands which determine the
electronic properties at low energies. Characteristic for the structure
of the $\pi$-band are the valence and conduction bands that touch
at the corner points of the first Brillouin zone, also called Dirac
points. Since
there is one $p_{z}$-electron per carbon atom, in isolated graphene
the valence band is completely filled whereas the conduction band
is empty. In the vicinity of the Dirac points the band structure exhibits
a linear dispersion relation, Fig. \ref{disps}a, resembling, up to a reduced propagation
velocity of $v=8.1\cdot10^{5}\mbox{ m/s}$, the one of massless relativistic
particles. From now on we focus on the region of linear dispersion.
A description in terms of a Dirac equation for the $p_{z}$-electrons
capturing the essential features of the $\pi$-band close to the Dirac
points can be obtained by a nearest neighbor tight binding calculation
~\cite{Castro-Neto08}.\\
The six corner points of the first Brillouin zone can be decomposed into
two subsets of equivalent Dirac points. As particular representatives
we choose \begin{equation}\vec{K}_{F}=F\frac{4\pi}{3\sqrt{3}a_{0}}\hat{k}_{x},\, F=\pm\label{eq:K-points}\end{equation}
where $a_{0}\approx0.14\mbox{ nm}$ is the nearest neighbor distance.
Restricting the discussion to the vicinity of $K_{\pm}$, the $\pi$-electrons are described by Bloch waves \begin{equation}
\varphi_{F\alpha}(\vec{r},\vec{\kappa})=\frac{1}{\sqrt{{2N}_{L}}}\sum_{p=\pm}\eta_{F\alpha\, p}(\vec{\kappa})\sum_{\vec{R}}e^{i\left(\vec{K}_{F}+\vec{\kappa}\right)\cdot\vec{R}}\chi_{\vec{R}\, p}(\vec{r}),\label{eq:bloch_waves}\end{equation}
 where $\vec{r},\vec{R}\in\mathbb{R}^2$, $N_{L}$ is the number of sites of the considered lattice,
$\alpha=\pm$ denotes the conduction and valence band, respectively,
and $\chi_{\vec{R}\, p}(\vec{r})$ is the $p_{z}$ orbital on sublattice
$p$ at lattice site $\vec{R}$. Finally, $\vec{\kappa}$ is the wave
vector relative to the Dirac point $\vec{K}_{F}$. Defining the spinors
\begin{equation}
\eta_{F\alpha}(\vec{\kappa}):=\left(\begin{array}{c}
\eta_{F\alpha\,-}(\kappa_x,\kappa_y)\\
\eta_{F\alpha\,+}(\kappa_x,\kappa_y)\end{array}\right),\ \mathrm{with}\ \kappa_{x}\equiv\hat{k}_{x}\cdot\vec{\kappa},\,\kappa_{y}\equiv\hat{k}_{y}\cdot\vec{\kappa} ,
\label{eq:f_bloch}\end{equation}
 it is found that they fulfill the Dirac equation\begin{equation}
\hbar v\left(-F\sigma_{x}\kappa_{x}+\sigma_{y}\kappa_{y}\right)\eta_{F\alpha}({\kappa}_x,\kappa_y)=\alpha\,\varepsilon({\kappa}_x,\kappa_y)\,\eta_{F\alpha}({\kappa}_x,\kappa_y),\label{eq:Diracequation}\end{equation}
 where \begin{equation}
\varepsilon({\kappa}_x,\kappa_y)=\hbar v\sqrt{\kappa_{x}^{2}+\kappa_{y}^{2}}\label{eq:disprel}\end{equation}
 reflects the linear dispersion relation and $\sigma_{x}$, $\sigma_{y}$
are the Pauli matrices. From (\ref{eq:Diracequation})
it follows that for $\kappa_y\neq\pm i \kappa_x$ it holds the relation\begin{equation}
\eta_{F\alpha\,+}(\kappa_x,\kappa_y)=-\alpha\frac{F\kappa_{x}-i\kappa_{y}}{\sqrt{\kappa_{x}^{2}+\kappa_{y}^{2}}}\,\eta_{F\alpha\,-}(\kappa_x,\kappa_y).\label{eq:relation_bloch}\end{equation}
 A specific solution of \Eq{eq:relation_bloch} we will use in
the remaining of the article is given by \[
\eta_{F\alpha\,-}({\kappa}_x,\kappa_y)=1,\qquad \eta_{F\alpha\,+}({\kappa}_x,\kappa_y)=-\alpha\frac{F\kappa_{x}-i\kappa_{y}}{\sqrt{\kappa_{x}^{2}+\kappa_{y}^{2}}}.\]

\subsection{Boundary effects in narrow ribbons}

So far no boundary effects have been included. However, we wish to
discuss the electronic properties of \textit{finite size} carbon nanoribbons,
which implies a wave function which vanishes all along the ribbon edges (open boundary conditions).
The geometry we want to study is depicted in Fig. \ref{graphene}. In particular we assume the long edges of the ribbon, along
the $y$ direction, in an armchair configuration, while the narrow terminations of the ribbon, along the $x$
axis, have zig-zag character. We are interested in quasi one-dimensional ribbons and thus we restrict our discussion to stripes with a large aspect ratio $L_{y}\gg L_{x}$
where $L_{x}$ and $L_{y}$ are the extensions in $x$ and $y$ direction, respectively.

From Fig. \ref{lattice} we can easily see that the two zig-zag ends each consist
of atoms either sitting on sublattice $p=+$ or $p=-$. We will use the convention that
the {}``left'' end at $y=0$ is formed by atoms living on sublattice
$p=-$ whereas on the other end (at $y=L_y$) we have atoms from sublattice $p=+$
only. To fulfill the appropriate boundary condition for the zig-zag ends, we have to construct new wave functions $\tilde{\varphi}_{F\alpha}(\vec{r},\vec{\kappa})$ from linear combinations of $\varphi_{F\alpha}(\vec{r},\vec{\kappa})$, in such a way that they vanish on the {}``missing''
atoms at the ends, namely on sublattice $p=+$ on the left end and
$p=-$ on the other end. A lengthy but straightforward calculation leads to
\numparts\begin{eqnarray}
&\tilde{\varphi}_{F\alpha}\left(\vec{r},(\kappa_x,\kappa_y )\right)\nonumber\\&=C_{zz}(F{\kappa}_x,\kappa_y)\left[\varphi_{F\alpha}\left(\vec{r}, (\kappa_x,\kappa_y ) \right)-\varphi_{F\alpha}\left(\vec{r}, (\kappa_x,-\kappa_y ) \right)\right],\label{eq:waveszig-zag}\end{eqnarray}
with a normalization constant\footnote{The normalization constant is non-trivial as the functions given in \Eq{eq:bloch_waves} are non-orthogonal. To verify that there is the dependence on $F\kappa_x$ and $\kappa_y$, but not on $\alpha$, \Eq{eq:bloch_waves} must be consulted in combination with \Eqs{eq:K-points}{eq:relation_bloch}.} $C_{zz}(F\kappa_x,\kappa_y)\in\mathbb{C}$ and the quantization condition~\cite{Brey06} \begin{equation}
e^{i 2\kappa_{y}L_{y}}=\frac{F\kappa_{x}+i\kappa_{y}}{F\kappa_{x}-i\kappa_{y}}.\label{eq:zig-zagquantcondition}\end{equation}\endnumparts

Additionally, the wave function we are looking for must also vanish at the armchair
edges. In contrast to the zig-zag end, the terminating
atoms where the wave function is required to go to zero are from both
sublattices. In order to build up suited linear combinations,
we have to mix states of Eq. (\ref{eq:waveszig-zag}) which belong to different
Dirac points. Then, the resulting wave function $\varphi_{\alpha}(\vec{r},\vec{\kappa})$ vanishes on the lattice sites with $R_{x}=0$ and $R_{x}=L_{x}$ for
\numparts\begin{eqnarray}
&\varphi_{\alpha}\left(\vec{r},(\kappa_x,\kappa_y )\right)\nonumber\\&=C_{ac}(\kappa_x,\kappa_y)\left[\tilde{\varphi}_{K_{+}\alpha}\left(\vec{r},(\kappa_x,\kappa_y )\right)-\tilde{\varphi}_{K_{-}\alpha}\left(\vec{r},(-\kappa_x,\kappa_y )\right)\right],\label{eq:wave_armchair}\end{eqnarray} with $C_{ac}(\kappa_x,\kappa_y)\in\mathbb{C}$ another normalization constant, and the quantization~\cite{Brey06}\begin{equation}
K_{\pm}\pm\kappa_{x}=\frac{\pi}{L_{x}}n_{x},\quad n_{x}\in\mathbb{Z}.\label{eq:armchairquantcondition}\end{equation}\endnumparts
Due to the relation $K_{\pm}=\pm K_0=\pm\frac{4\pi}{3\sqrt{3}a_0}$, the two conditions in \Eq{eq:armchairquantcondition} are equivalent.

\subsection{The eigenstates of metallic ACNs}

In total we obtain with Eqs. (\ref{eq:waveszig-zag}) and (\ref{eq:wave_armchair})
the following expression for the eigenstates of noninteracting electrons
in finite size ACNs,\begin{equation}
\varphi_{\alpha}\left(\vec{r},(\kappa_x,\kappa_y)\right)=C(\kappa_x,\kappa_y)\sum_{F,r=\pm}(Fr)\,\varphi_{F\alpha}\left(\vec{r},(F\kappa_{x},r\kappa_{y})\right),\label{eq:eigenstates}\end{equation}
 where the generally complex number $C(\kappa_x,\kappa_y)=C_{zz}(\kappa_x,\kappa_y)C_{ac}(\kappa_x,\kappa_y)$ guarantees that $\varphi_{\alpha}(\vec{r},(\kappa_x,\kappa_y))$ is normalized to
$1$.\\

We want to investigate now the solutions of the Dirac equation fulfilling our quantization conditions Eqs. (\ref{eq:zig-zagquantcondition}) and (\ref{eq:armchairquantcondition}). It has been shown~\cite{Fujita96,Nakada96} that the presence of
zig-zag ends leads to the formation of localized states characterized
by a purely imaginary $\kappa_{y}$ giving rise to an enhanced density
of states around the Dirac energy. 
Though those states are localized at the zig-zag ends with an exponential decay in $y$ direction, they are  of decisive
relevance for the transport properties of ACNs.\\

At first, however, let us focus on the {\bf\emph{bulk states}}, where both $\kappa_{x}$ and $\kappa_{y}$ are real numbers.
\begin{figure}
\begin{center}\includegraphics[width=0.98\columnwidth]{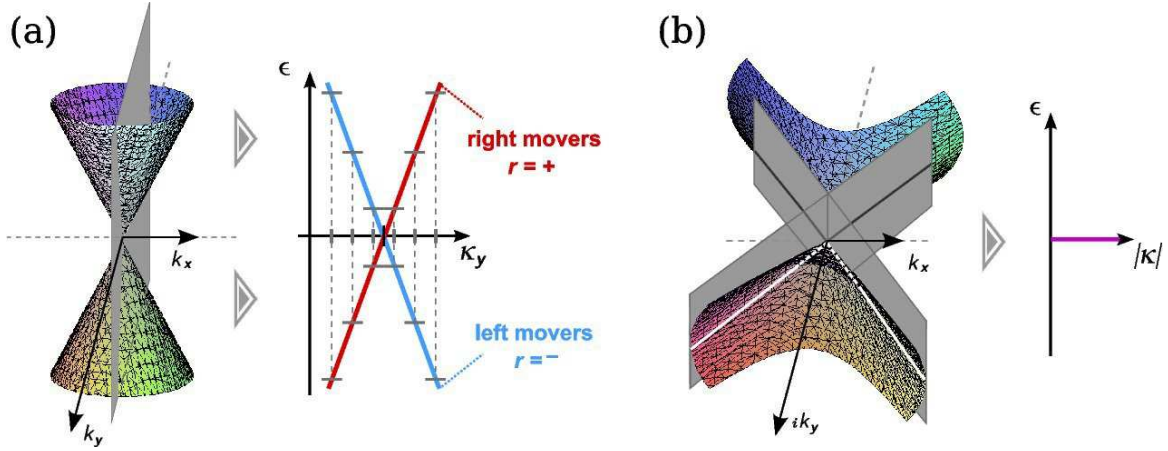}\end{center}
\caption{The dispersion relation of graphene, \Eq{eq:disprel}, for (a) real and (b) imaginary momenta $\kappa_y$. In both cases the boundary condition at the armchair edge mixes wave functions belonging to both Dirac points. (a) For solutions with real $\kappa_y$, in the low energy regime only sub-bands with $\kappa_x=0$ play a role due to the condition $L_x\ll L_y$. The corresponding dispersion relation emerges thus from the intersection of the Dirac cone with the plane $k_x=0$. (b) For solutions with imaginary $\kappa_y$ the cone opens along $i\hat{k}_y$ with a slope determined by $\hbar v$. As for the eigenstates it must hold $\kappa_y=\pm i\kappa_x$, the corresponding dispersion relation is obtained from the intersection of the cone with the two planes $k_x=\pm i k_y$. There is only tangency along two straight lines within the $k_x$-$i k_y$-plane, resulting in a dispersion which is identically zero.}\label{disps}
\end{figure}
Since $L_{y}\gg L_{x}$, the quantization condition \Eq{eq:armchairquantcondition}
leads to the formation of sub-bands assigned to different $\kappa_{x}.$ For a metallic ACN, the sub-band must cross the Dirac points $F=K_\pm$, which demands $\kappa_{x}=0$, or equivalently $n_{x}=\frac{L_{x}}{\pi}K_{0},\: n_{x}\in\mathbb{N}$, which follows from 
\Eq{eq:armchairquantcondition}. The width $L_{x}$ depends on the number
$M$ of hexagons in a row parallel to $\hat{x}$ like $L_{x}=\sqrt{3}a_{0}(M+1/2)$ (compare Fig. \ref{graphene}).
Hence $n_{x}=\frac{1}{3}(4M+2)$ (cf. \Eq{eq:K-points}). Obviously, this means
that the geometrical condition for having metallic ACNs with gap-less sub-bands
reads $M\,\mathrm{mod}\, 3=1$.
Focussing on such ribbons, $\vec{\kappa}\propto\hat{k}_y$ as $\kappa_x=0$, which corresponds to a cut through the Dirac cone as shown in \Fig{disps}a. The corresponding states are characterized by the band index $\alpha$
and $\kappa_{y}$, see \Eq{eq:wave_armchair}. With $\kappa_{x}=0$, Eq. (\ref{eq:zig-zagquantcondition}) yields as allowed values of $\kappa_{y}$:\begin{equation}\label{kappay}
\kappa_{y}=\frac{\pi}{L_{y}}\left(n+\frac{1}{2}\right),\, n\in\mathbb{Z}.\end{equation}Since by definition, Eq. (\ref{eq:eigenstates}),
$\varphi_{\alpha}\left(\vec{r},(0,\kappa_{y})\right)=-\varphi_{\alpha}\left(\vec{r},(0,-\kappa_{y})\right)$, we can further restrict for each $\alpha$ our analysis to either $\kappa_y>0$ or $\kappa_y<0$.
Thus it is an allowed choice to just consider states with $\sgn(\kappa_y)=\sgn(\alpha)$, which we define as \[
\varphi_{\kappa_y}^{{ \bulk}}(\vec{r}):=\varphi_{\alpha=\sgn(\kappa_{y})}\left(\vec{r},(0,\kappa_y)\right).\]
Doing so, we select the positive slope of the two branches of the dispersion relation Fig \ref{disps}a.
Bearing in mind the form of the graphene Bloch waves, \Eq{eq:bloch_waves},
we can of course express the states $\varphi^{{ \bulk}}_{\kappa_{y}}$ in terms
of the sublattice wave functions $\varphi_{Fp}:=\left({2N}_{L}\right)^{-1/2}\sum_{\vec{R}}e^{i\left(\vec{K}_{F}+\vec{\kappa}\right)\cdot\vec{R}}\chi_{\vec{R}\, p}(\vec{r})$,\[
\varphi^{{ \bulk}}_{\kappa_{y}}(\vec{r})=\frac{1}{2}\sum_{Fpr}f_{Fpr}\varphi_{Fp}\left(\vec{r},(0,r\kappa_y)\right),\]
 where up to a complex prefactor the coefficients $f_{Fpr}$ are given
by\begin{equation}
f_{F+r}=rF\,,\quad f_{F-r}=i F.
\label{eq:f_Fpr}\end{equation}
 Note that the index $r$ here denotes right ($r=+$) and left ($r=-$)
moving waves [compare also to \Fig{disps}a].\\


Now we turn to the {\bf\emph{end states}}, emerging for purely imaginary $\kappa_{y}$~\cite{Castro-Neto08}, which are allowed by
both the Dirac equation \Eq{eq:Diracequation} and the quantization condition \Eq{eq:zig-zagquantcondition}.
In more detail, there exist two imaginary solutions for
each $\kappa_{x}>{1}/{L_{y}}$, which holds in ACNs for all $\kappa_{x}=n{\pi}/{L_{x}},\, n\in\mathbb{N}$.
Besides, the relation $L_{x}\ll L_{y}$ causes that to a very good approximation 
$\kappa_{y}=\pm i\kappa_{x}$ 
satisfies Eq. (\ref{eq:Diracequation}) and Eq. (\ref{eq:zig-zagquantcondition}).
The corresponding dispersion relation is given in Fig. \ref{disps}b.
Notice that \Eq{eq:relation_bloch} is not applicable to $\kappa_{y}=\pm i\kappa_{x}$ (as explicitly exempted before), and instead the spinors fulfilling the Dirac equation \Eq{eq:Diracequation} are given by
$\eta_{F\alpha p}(\kappa_x,\pm i\kappa_x)=\delta_{p,\pm F}$. 
Using this in Eq. (\ref{eq:bloch_waves}) and following the steps leading to \Eq{eq:eigenstates}, we obtain instead after straightforward insertions
\begin{eqnarray*}\varphi_{\alpha}(\vec{r},(\kappa_x,\pm i\kappa_{x}))=\pm C(\kappa_x,\pm i\kappa_{x})\sum_{Fp} Fp\, \varphi_{Fp}\left(\vec{r},(F\kappa_x,i p\kappa_{x})\right).\end{eqnarray*}
 The corresponding ACN eigenstates can be chosen such that they live
on one sublattice $p=\pm$ only: \[
\varphi_{p \kappa_{x}}^{{ \edge}}(\vec{r})=\tilde{C}(\kappa_x)\sum_{F}F\varphi_{Fp}\left(\vec{r},(F\kappa_x,i p\kappa_x)\right),\]
 where $\tilde{C}(\kappa_x)$ is a normalization constant.
It is evident that the decay length
of $\varphi_{p\kappa_{x}}^{{ \edge}}(\vec{r})$ from one of the ends to
the interior of a specific ACN is $L_{x}/({n_{x}\pi})$, which is
always much shorter than the length of the ribbon. That is, one finds
\textit{localized end states}.

  From the dispersion relation Eq. (\ref{eq:disprel}) it is easy to see that the energy of the
end states is zero. Consequently, they will be unpopulated below half filling,
but as soon as the Dirac point is reached, one electron will get trapped at each end (in an interacting system, Coulomb repulsion will hinder a second electron to enter).
So we can conclude that at energies around the Dirac energy, not only
the extended states with $\kappa_{x}=0$, but also the localized
end states can be of importance.

\subsection{Electron and Hamilton operator of the metallic ACN}
All in all, the appropriate
operator describing an electron with spin $\sigma$ at position $\vec{r}$
reads in the low energy regime\begin{equation}
\hPsi_{\sigma}(\vec{r})=\sum_{\kappa_{y}=(\mathbb{Z}+0.5)\pi/L_y}\varphi^{{ \bulk}}_{\kappa_{y}}(\vec{r})\hc_{\sigma \kappa_{y}}+\sum_{p}\sum_{\kappa_{x}=\mathbb{N}\pi/L_x}\varphi_{p\kappa_{x}}^{{ \edge}}(\vec{r})\hd_{\sigma p\kappa_{x}},\label{elops}\end{equation}
where $\hc_{\sigma k},\ \hd_{\sigma p k}$ are the annihilation operators for electrons of momentum $k$ and spin $\sigma$ in the bulk or end states, respectively.
 The one-dimensional (1D) character of ACNs at low energies becomes evident by defining
the slowly varying electron operators \begin{equation}
\hpsi_{r\sigma}(y):=\sum_{\kappa_{y}=(\mathbb{Z}+0.5)\pi/L_y}\varphi^{{ \bulk}}_{\kappa_{y}}(\vec{r})\hc_{\sigma \kappa_{y}}=\frac{1}{\sqrt{2L_y}}\sum_{\kappa_{y}=(\mathbb{Z}+0.5)\pi/L_y}e^{i r\kappa_y}\hc_{\sigma \kappa_{y}}\label{eq:1Delop}\end{equation}
 such that we obtain \begin{equation}
\hpsi_{\sigma}(\vec{r})=\sqrt{L_y/2}\sum_{Fpr}f_{Fpr}\varphi_{Fp}(\vec{r})\hat{\psi}_{r\sigma}(y),\label{eq:3Dto1D}\end{equation}
 with $\varphi_{Fp}(\vec{r}):=\varphi_{Fp}(\vec{r},\vec{\kappa}=(0,0))$.\\

From the dispersion relations \Fig{disps}, it is easy to give the Hamilton operator of the noninteracting metallic ACN,
\begin{equation}\hH_{0}=\hbar v\sum_{\sigma\kappa_{y}}\kappa_{y}\hc_{{\sigma\kappa}_{y}}^{\dagger}\hc_{{\sigma\kappa}_{y}},\label{ACN_ham}\end{equation}
with the Fermi velocity $v=8.1\times10^{5}\mbox{ m/s}$ corresponding to the absolute value of the slopes of the linear branches in \Fig{disps}a. There is no contribution of the end states as those have zero energy, see \Fig{disps}b. With the allowed values for $\kappa_y$, \Eq{kappay}, there results obviously a level spacing \begin{equation}\label{eps0}\varepsilon_0:={\hbar\pi v}/{L_y}.\end{equation}

\section{The interaction Hamiltonian\label{sec:interaction-ham}}

This section is dedicated to the determination of the eigenstates of an interacting ACN.
The cornerstones of the derivation and an analysis of the resulting spectrum were already
presented in our recent short manuscript \cite{Koller09}. We provide here merely a
compact outline of the technical steps and refer the interested readers for details to
the appendices. A discussion of the main features needed to understand the transport results
is provided in \Sct{sec:minmod}.\\

In the following we concentrate both on interaction effects regarding
the extended bulk states in ACNs as well as on correlations between end and bulk
states.
Ignoring exchange effects, the many-body end states can be
spin- or edge- degenerate.
Above half filling, within a reasonable energy range, both end states can be assumed to be populated with one single
electron only. This is because their charging energy exceeds, due to the strong localization in position space,
the charging energy of the extended states by far: A simple estimation modelling
the Coulomb repulsion between two electrons localized at the same ribbon end via the
Ohno potential given below, \Eq{Ohno}, yields energies of the order of $0.1\,$eV for ribbon width
around $10\,$nm. In contrast, typical charging energies for the bulk states of such ACNs range around $1-10\,$meV.
The only relevant scattering processes between bulk and end states
mediated by the Coulomb interaction are thus of the form \begin{eqnarray}
\nonumber\hV_{{\eb}}=\sum_{\sigma\sigma'}\sum_{\tilde{p}}\int\!\!\int \rmd^{3}r\,\rmd^{3}r'&\left( \hpsi_{\sigma}^{\dagger}(\vec{r})\hpsi_{\sigma'\tilde{p}}^{{ \edge}\dagger}(\vec{r}\,') \right.U(\vec{r}-\vec{r}\,')\hpsi_{\sigma'\tilde{p}}^{{ \edge}}(\vec{r}\,')\hpsi_{\sigma}(\vec{r})\\&
+\hpsi_{\sigma}^{\dagger}(\vec{r}) \hpsi_{\sigma'\tilde{p}}^{{ \edge}\dagger}(\vec{r}\,')\left.U(\vec{r}-\vec{r}\,')\hpsi_{\sigma'}(\vec{r}\,')\hpsi_{\sigma \tilde{p}}^{{ \edge}}(\vec{r})\right).\label{V-edge-bulk}\end{eqnarray}
All other processes should be strongly suppressed for energy reasons.
For the bulk-bulk interaction, our scattering potential is described by the expression
\begin{equation}
\hV_{{\bb}}=\frac{1}{2}\sum_{\sigma\sigma'}\int \!\! \int \rmd^{3}r\, \rmd^{3}r'\,\hpsi_{\sigma}^{\dagger}(\vec{r})\hpsi_{\sigma'}^{\dagger}(\vec{r}\,')U(\vec{r}-\vec{r}\,')\hpsi_{\sigma}(\vec{r})\hpsi_{\sigma'}(\vec{r}\,').\label{V-bulk-bulk}\end{equation}
In both Eqs. (\ref{V-edge-bulk}) and (\ref{V-bulk-bulk}), the function $U(\vec{r}-\vec{r}\,')$ models the screened 3D Coulomb interaction
potential. For our calculations we use the Ohno potential~\cite{Barford05},
\begin{equation}\label{Ohno}
U(\vec{r}-\vec{r}\,')=U_0 \left(1+\left(\frac{U_0\epsilon|\vec{r}-\vec{r}\,'|}{14.397[\mathrm{\AA}\,\mathrm{eV}]}\right)^2\right)^{-\frac{1}{2}},
\end{equation}
\noindent with $U_0=15\ \mbox{eV}$ \cite{Fulde95} and $\epsilon\simeq1.4-2.4$~\cite{Egger97} the dielectric constant of graphene.\\
We proceed now with a discussion of the two different expressions. We start with the bulk-bulk processes, where the analysis follows
largely the lines of an earlier work on interaction effects in metallic single wall carbon nanotubes (SWCNTs)~\cite{Mayrhofer08}.

\subsection{Bulk-bulk interaction}

 With the help of the reformulation of the 3D electron operator in
terms of the 1D operators $\hpsi_{r\sigma}(y)$, Eq. (\ref{eq:3Dto1D}),
we obtain after integrating over the coordinates perpendicular to
the ribbon axis an effectively 1D expression for the interaction,
\begin{equation}
\hV_{{\bb}}=\sum_{\{[r]\}}\sum_{\sigma\sigma'}\int\!\!\int \rmd y\,\rmd y'\,\hpsi_{r_{1}\sigma}^{\dagger}(y)\hpsi_{r_{2}\sigma'}^{\dagger}(y')\frac{1}{2}U_{[r]}(y,y')\hpsi_{r_{3}\sigma'}(y')\hpsi_{r_{4}\sigma}(y),\label{eq:V_1}\end{equation}
 where $\sum_{\{[I]\}}$ denotes the sum over all possible quadruples
$[I_{1},I_{2},I_{3},I_{4}]$, in the former case for the band index $I=r$. The spin-independent 1D interaction potential $U_{[r]}(y,y')$
reads\begin{eqnarray}\nonumber
&U_{[r]}(y,y')=\frac{L^2_y}{4}\sum_{pp'}\sum_{\{[F]\}}f_{F_{1}pr_{1}}^{*}f_{F_{2}p'r_{2}}^{*}f_{F_{3}p'r_{3}}f_{F_{4}pr_{4}}\\
&\times\int_{\bot}\!\int_{\bot}\rmd^{2}r\,\rmd^{2}r'\,\varphi_{F_{1}p}(\vec{r})\varphi_{F_{2}p'}(\vec{r}\,')U(\vec{r}-\vec{r}\,')\varphi_{F_{3}p'}(\vec{r}\,')\varphi_{F_{4}p}(\vec{r}),\label{eq:U1Ddef}\end{eqnarray}
with $\int_\bot$ indicating that the integration has to be performed over the coordinates perpendicular to $y,y'$ (i.e. $x,x',z,z'$).
Exploiting the explicit form, Eq. (\ref{eq:f_Fpr}), of the coefficients $f_{Fpr}$ we can easily identify
those scattering processes which are indeed mediated by $U_{[r]}(y,y')$.
In detail, \begin{equation}
U_{[r]}(y,y')=\frac{1+r_{1}r_{2}r_{3}r_{4}}{2}\,U^{{ intra}}(y,y')+\frac{r_{1}r_{4}+r_{2}r_{3}}{2}\,U^{{ inter}}(y,y'),\label{eq:UtoUinterintra}\end{equation}
where we have defined the potentials \begin{eqnarray*}
&U^{{ intra/inter}}(y,y'):=\frac{L_y^2}{2}\sum_p\sum_{\{[F]\}}F_{1}F_{2}F_{3}F_{4}\\&
\times\int_{\bot}\!\int_{\bot}\rmd^{2}r\,\rmd^{2}r'\,\varphi^*_{F_{1}p}(\vec{r})\varphi^*_{F_{2}\pm p}(\vec{r}\,')U(\vec{r}-\vec{r}\,')\varphi_{F_{3}\pm p}(\vec{r}\,')\varphi_{F_{4}p}(\vec{r}),\end{eqnarray*}
 describing interactions between electrons residing on the same/different
sublattices. From Eq. (\ref{eq:UtoUinterintra}) it is clear that a non-vanishing
interaction potential can only be assigned to processes characterized
by $r_{1}r_{4}=r_{2}r_{3}$, and those are the forward ($f$)-, back ($b$)-, and umklapp ($u$)- scattering. Denoting
the scattering type by $S_{I}$, the corresponding quadruples are $[I]_{S_{I}=f^{\pm}}=[I,\pm I,\pm I,I]$,
$[I]_{b}=[I,-I,I,-I]$ and $[I]_{u}=[I,I,-I,-I]$. In total this
means that we can rewrite \Eq{eq:V_1} as\footnote{The spin quadruple $[\sigma,\sigma',\sigma',\sigma]$ yields possible configurations $[\sigma]_{f^\pm}=[\sigma,\pm\sigma,\pm\sigma,\sigma]$.}\[
\hV_{{\bb}}=\sum_{S_{r}=u,b,f^{\pm}}\sum_{S_{\sigma}=f^{\pm}}\hV^{{\bb}}_{S_{r}S_{\sigma}}.\]
In the case of umklapp- and back-scattering with respect to $r$, the
potential $U_{[r]}(y,y')$ is proportional to the difference of the
inter- and intra-lattice interaction potential. Since the latter potentials
differ only on the length scale of the lattice spacing \mbox{$a_{0}\approx0.14\,$nm}, this means that in the case of $S_{r}=u,b$ the effective 1D potential
$U_{[r]}(y,y')$ can be considered as point-like. Introducing the coupling
constant \[
u:=\frac{1}{4L_y^2}\int\!\!\int \rmd y\,\rmd y'\,\left(U^{{ intra}}(y,y')- U^{{ inter}}(y,y')\right),\]
we can set in good approximation $U_{[r]}(y,y')=u\,\delta(y-y')$ and write the short-ranged interaction processes as \numparts\begin{eqnarray}
\hV^{{\bb}}_{bf^{\pm}}&=&\frac{u}{2}\sum_{r\sigma}\int \rmd y\,\hpsi_{r\sigma}^{\dagger}(y)\hpsi_{-r\pm\sigma}^{\dagger}(y)\hpsi_{r\pm\sigma}(y)\hpsi_{-r\sigma}(y),\label{V_b-b_shortranged-b}\\
\hV^{{\bb}}_{uf^{-}}&=&\frac{u}{2}\sum_{r\sigma}\int \rmd y\,\hpsi_{r\sigma}^{\dagger}(y)\hpsi_{r-\sigma}^{\dagger}(y)\hpsi_{-r-\sigma}(y)\hpsi_{-r\sigma}(y).\label{V_b-b_shortranged-u}\end{eqnarray}\endnumparts
Since $u$ is derived from a short-ranged interaction it scales
inversely with the size of the underlying ribbon. We find typical values~\cite{DiEl} of $uL_x/\varepsilon_0 = 0.1\,$nm for a level spacing $\varepsilon_0$, \Eq{eps0}. The
process $V^{{\bb}}_{uf^{+}}$ vanishes identically, because it involves the operator product $\hpsi_{-r\sigma}(y)\hpsi_{-r\sigma}(y)=0$. Only the forward scattering
processes $V^{{\bb}}_{f^{\pm}f^{\pm}}$ are long ranged.

Since easily diagonalizable by bosonization, it is convenient to identify the density-density processes among the relevant bulk-bulk interaction processes,
such that $\hV_{{\bb}}$ can be decomposed into 
$\hV_{{\bb}}=\hV^{{\bb}}_{\rho\rho}+\hV^{{\bb}}_{n\rho\rho},$ 
with the density-density and non-density-density parts given by
\[
\hV^{{\bb}}_{\rho\rho}=\hV^{{\bb}}_{f^{\pm}f^{\pm}}+\hV^{{\bb}}_{bf^{+}} \quad
\mathrm{and}\quad
\hV^{{\bb}}_{n\rho\rho}=\hV^{{\bb}}_{bf^{-}}+\hV^{{\bb}}_{uf^{-}}, \]
respectively.

\subsection{End-bulk interaction}
\newcommand{\kappax}{}
In ACNs, we additionally have to consider scattering between the electrons living in the bulk of the ribbon and the electrons trapped in the end state existing at both zig-zag terminations of the stripe. Below half-filling, the end states are unpopulated and thus all terms discussed in the following would be zero a priori. The range we want to concentrate on is the low energy regime above half-filling, where exactly one electron will permanently occupy each end state, so that we have in total two end electrons interacting with our bulk electrons.\\
We can certainly impose that the wave functions of the localized $p_z$-orbitals have non-vanishing overlap for electrons on the same sublattice only. Moreover we demand that both end electron operators have to belong to the same end, and thus to the same sublattice, in order to give a nonzero contribution\footnote{Operators acting on different ends would change the population of each end state, which is not allowed.}.
If we insert then  into Eq. (\ref{V-edge-bulk}) the decomposition Eq. (\ref{eq:3Dto1D}) for the bulk electron operator and $\psi^{{\small \edge}}_{\tilde{p}\sigma}(\vec{r})=\sum_{\kappa_x}\varphi_{\tilde{p}\kappa_{x}}^{{ \edge}}(\vec{r})\hd_{\sigma \tilde{p}\kappa_{x}}$ for the end electron operator, and set in the coefficients from \Eq{eq:f_Fpr}, we obtain:
\begin{eqnarray}
&\hV_{{\eb}}=\frac{L_y}{2}\sum_{\kappa_x}\sum_{\sigma\sigma'}\sum_{\tilde{p}}\sum_{rr'}\int\!\!\int \rmd y\, \rmd y'\ \hd_{\sigma'\kappa_x\tilde{p}}^{\dagger}\hpsi_{r\sigma}^{\dagger}(y)\nonumber\\
&\times\left(\delta_{r,r'}\hpsi_{r'\sigma}(y)\,U_\rho^{\kappa_x\tilde{p}}(y,y')\hd_{\sigma'\tilde{p}\kappa_x}-(rr')^{\delta_{\tilde{p},+}}\hpsi_{r'\sigma'}(y')\,U_{n\rho}^{\kappa_x\tilde{p}}(y,y')\hd_{\sigma \tilde{p}\kappa_x}\right).\label{V_e-b}
\end{eqnarray}
Thereby, symmetry arguments have lead to the demand $r=r'$ for the part containing the interaction potential related to densities,
\begin{eqnarray*}
&U_{\rho}^{\kappa_x\tilde{p}}(y,y')=\sum_{FF'}\int_{\bot}\!\int_{\bot}\rmd^{2}r\,\rmd^{2}r'\,FF'\,U(\vec{r}-\vec{r}\,')\\&\times\left(\varphi^*_{F+}(\vec{r})\varphi_{F'+}(\vec{r})+\varphi^*_{F-}(\vec{r})\varphi_{F'-}(\vec{r})\right)\left|\varphi_{\tilde{p}\kappa_x}^{{ \edge}}(\vec{r}\,')\right|^2,
\end{eqnarray*}
which, again for symmetry reasons, fulfils further
\[\int\rmd y' U_\rho^{\kappa_x+}(y,y')=\int\rmd y' U_\rho^{\kappa_x-}(y,y')\equiv t^{\kappa_x}_\rho(y)\approx t_\rho(y)\ \forall\kappa_x.\]
We define a density-density part of the end-bulk interaction correspondingly as
\begin{equation}\label{V_e-b_rr}\hV^{{\eb}}_{\rho\rho}={L_y}\int \rmd y\ t_\rho(y)\left(\sum_{r\sigma}\hpsi_{r\sigma}^{\dagger}(y)\hpsi_{r\sigma}(y)\right),
\end{equation}
where we exploited that each end state is populated with exactly one electron.\\
The second potential,
\begin{eqnarray*}
U_{n\rho}^{\kappa_x\tilde{p}}(y,y')=&\sum_{FF'}\int_{\bot}\!\int_{\bot}\rmd^{2}r\,\rmd^{2}r'\,FF'\,U(\vec{r}-\vec{r}\,')\\&\times\varphi^*_{F\tilde{p}}(\vec{r})\varphi_{F'\tilde{p}}(\vec{r}\,')\varphi_{\tilde{p}\kappa_x}^{{ \edge}*}(\vec{r}\,')\varphi_{\tilde{p}\kappa_x}^{{ \edge}}(\vec{r}),
\end{eqnarray*}
can be considered point-like due to the localization of the end states at $y_{\tilde{p}=-}=0$ or  $y_{\tilde{p}=+}=L_y$, and hence simplifies to
\begin{equation*}
U_{n\rho}^{\kappa_x\tilde{p}}(y,y')=\left(\delta_{\tilde{p},-}\delta(y-0)\delta(y'-0)+\delta_{\tilde{p},+}\delta(y-L_y)\delta(y'-L_y)\right)t^{\kappa_x}_{n\rho},\end{equation*}
with a short-range coupling constant which is independent of $\tilde{p}$ for symmetry reasons,
\begin{equation*}\label{eq:t}
t^{\kappa_x}_{n\rho}:=\sum_{FF'}\int\!\!\int\rmd^{3}r\,\rmd^{3}r'\,FF'\,U(\vec{r}-\vec{r}\,')\varphi^*_{F+}(\vec{r})\varphi_{F'+}(\vec{r}\,')\varphi_{+\kappa_x}^{{ \edge}*}(\vec{r}\,')\varphi_{+\kappa_x}^{{ \edge}}(\vec{r}).
\end{equation*}
%
For an ACN of width $L_x$ ranging from $5$ to $25$\,nm, one finds~\cite{DiEl} $t^{\kappa_x}_{n\rho} \approx t_{n\rho}=:t$, with $tL_x/\varepsilon_0 \approx 0.55\,$nm, practically independent of $\kappa_x$.
This means that the short-ranged end-bulk scattering is comparable in strength to the exchange interactions induced by
the bipartite sublattice structure, and consequently we have to account for a non-negligible contribution
\begin{equation}\hV^{{\eb}}_{n\rho\rho}=-\frac{L_y}{2}\sum_{\sigma\sigma'rr'p}\left(\delta_{{p},+}rr'+\delta_{{p},-}\right)
\hpsi_{r\sigma}^{\dagger}(y_{{p}})\hpsi_{r'\sigma'}(y_{{p}})\,t\sum_{\kappa_x}\hd_{\sigma'{p}\kappa_x}^{\dagger}\hd_{\sigma {p}\kappa_x}
\label{V_e-b_nrr}\end{equation}
in the total end-bulk interaction potential $\hV_{{\eb}}=\hV^{{\eb}}_{\rho\rho}+\hV^{{\eb}}_{n\rho\rho}$.

%

\subsection{Diagonalization of the ACN Hamiltonian\label{sec:diag}}

We can diagonalize the Hamiltonian $\hH_{0}+\hV^{{\bb}}_{\rho\rho}+\hV^{{\eb}}_{\rho\rho}$ by bosonization
and subsequently express the total ACN Hamiltonian, \begin{equation}\hH_\odot:=\hH_{0}+
\underbrace{\hV^{{\bb}}_{\rho\rho}+\hV^{{\eb}}_{\rho\rho}}_{=:\hV_{\rho\rho}}+\underbrace{\hV^{{\bb}}_{n\rho\rho}+\hV^{{\eb}}_{n\rho\rho}}_{=:\hV_{n\rho\rho}},\label{ACN-ham}\end{equation} in the eigenbasis
of $\hH_{0}+\hV_{\rho\rho}$. 
A numerical diagonalization of the so constructed total Hamiltonian, however, yields reliable results only away from half-filling: as the eigenbasis needs to be truncated
for the calculation, it is crucial that $\hV_{n\rho\rho}$ is just a perturbation to  $\hH_{0}+\hV_{\rho\rho}$ in the sense that it only mixes states close in energy to each other.
In the direct vicinity of the Dirac points, the process $\hV^{{\bb}}_{uf^{-}}$ breaks this demand, while it vanishes away from half-filling, as it will become clear in the course of this section.

\paragraph{Diagonalization of the density-density part\label{sec:diagdiag}}

Diagonalization of $\hH_{0}+\hV_{\rho\rho}$ can be achieved by bosonization.
The end result of the procedure on which more details are given in \App{bosonization}
is\begin{equation}
\hH_{0}+\hV_{\rho\rho}=\frac{1}{2}\left(W_{0}\hat{\mathcal{N}}_{c}^{2}
+\varepsilon_{0}\hat{\mathcal{N}}_c+\left\{\varepsilon_{0}-\frac{u}{2}\right\}\sum_\sigma\hat{\mathcal{N}}^2_{\sigma}\right)+\sum_{j,q>0}\varepsilon_{j q}\,\ha_{j q}^{\dagger}\ha_{j q}.\ \label{eq:H0Vrr_diag}\end{equation}
The first three purely fermionic contributions in \Eq{eq:H0Vrr_diag} account for charging and shell filling.  The  last term counts the bosonic excitations of the system, created/annihilated
by the operators $\ha_{j q}^{\dagger}$ / $\ha_{j q}$. The two
 channels $j=c,s$ are associated to charge $(c)$ and spin $(s)$ excitations.
 The excitation energies are 
\[
\varepsilon_{jq}=\sqrt{X_{jq}^{2}-A_{jq}^{2}}\,,\]\begin{equation}\mathrm{with}\ \
X_{jq}  =  n_{q}\left[\delta_{j,c}W_{q}+\varepsilon_{0}\right]\,,\ A_{jq}  =  n_{q}\left[\delta_{j,c}W_{q}-\frac{u}{2}\right].\label{eq:X_xq}\end{equation}
The energies of the charge channel are dominated by the long-ranged interactions via
the coefficients
\begin{equation}\label{Wq}
W_{q}=\frac{1}{L_y^{2}}\int_0^{L_y}\!\!\! \rmd y\,\int_0^{L_y}\!\!\!\rmd y'\left(U^{{ intra}}(y,y')+U^{{ inter}}(y,y')\right)\cos(qy)\cos(qy').
\end{equation}

Finally, we can give the eigenbasis of $H_{0}+V^{{\bb}}_{\rho\rho}$ in terms of states \begin{equation}
\left|\vec{N},\vec{\sigma}^{{ \edge}},\vec{m}\right\rangle =\prod_{j=c,s}\prod_{q>0}{\left(a_{jq}^{\dagger}\right)^{m_{jq}}}/{\sqrt{m_{jq}!}}\left|\vec{N},\vec{\sigma}^{{ \edge}},0\right\rangle ,\label{eq:eigenstates_H0Vrr}\end{equation}
 where $|\vec{N},\vec{\sigma}^{{ \edge}},0\rangle $ has no bosonic excitation.
The fermionic configuration $\vec{N}=(N_{\uparrow},N_{\downarrow})$ defines the number of electrons in each spin band. The occupation of the end states is determined by $\vec{\sigma}^{{ \edge}}=(\sigma^{{ \edge}}_{+},\sigma^{{ \edge}}_{-})$, where `$-$' relates to $y_-=0$ and `$+$' to $y_+=L_y$ as before. Below half filling the end states are empty, such that there is only one possible configuration: $\sigma^{{ \edge}}_+=0=\sigma^{{ \edge}}_-$. Above half filling, exactly one electron occupies each end state and thus $\sigma^{{ \edge}}_+,\sigma^{{ \edge}}_-\in\{\uparrow,\downarrow\}$. Finally, $\vec{m}=(\vec{m}_s,\vec{m}_c)$, with $m_{jq}=(\vec{m}_j)_q$ containing the information how many bosonic excitations are present in level $q$ for mode $j=c,s$.\\

\paragraph{Non-density-density interaction\label{sub:The-matrix-elements}}
In the following we use the states from Eq. (\ref{eq:eigenstates_H0Vrr})
as basis to examine the effect of $\hV_{n\rho\rho}$. For this
purpose we evaluate the matrix elements of the
potentials $\hV^{{\bb}}_{n\rho\rho}$ and $\hV^{{\eb}}_{n\rho\rho}$,
using the bosonization identity for the 1D electron operators.

Generally, $\hV^{{\bb}}_{n\rho\rho}$ does not conserve the quantity $\vec{m}$, while it must neither mix states with different electron configurations $\vec{N}$, nor with different end spin configurations $\vec{\sigma}^{{ \edge}}$: the Coulomb interaction between bulk electrons cannot change the quantity $S_z=\frac{1}{2}(N_\uparrow-N_\downarrow)$, and it cannot touch the end states. Further, we already know that both processes $\hV^{{\bb}}_{uf^-}$ and $\hV^{{\bb}}_{bf^-}$
contained in $\hV^{{\bb}}_{n\rho\rho}$ are effectively local interactions, see \Eqs{V_b-b_shortranged-b}{V_b-b_shortranged-u}, such that the matrix elements of the non-diagonal bulk-bulk interaction scale with the exchange-coupling parameter $u$,
\begin{eqnarray}
&\left\langle \vec{N}\vec{\sigma}^{{ \edge}}\vec{m}\left|\hV^{{\bb}}_{n\rho\rho}\right|\vec{N}'\vec{\sigma}^{{ \edge}}\vec{'m}'\right\rangle=\frac{u}{2L_y} \delta_{\vec{N},\vec{N}'}\delta_{\vec{m}_c,\vec{m}'_c}\delta_{\vec{\sigma}^{\edge},\vec{\sigma}^{\edge}{}'}\nonumber\\&\times\sum_{r\sigma}\int \rmd y\frac{e^{-i \frac{2\pi}{L_y}\sgn(r)\left(N_\sigma-N_{-\sigma}\right)y}}{4\sin^{2}\left(\pi y/{L_y}\right)}\prod_{q}F(\tilde{\lambda}_{[r]_b[\sigma]_{f^-}}^{s q}(y),m_{s q},m'_{s q}).\label{eq:ME_V_nrr}\end{eqnarray}
The derivation of this expression is given in \App{app:matrixelements-bulk}, as well as the definitions of $F(\lambda,m,m')$ and $\tilde{\lambda}_{[r][\sigma]}^{jq}(y)$, \Eq{eq:Fvonlambda1} and \Eq{eq:lambda_tilde}, respectively. In fact it turns out that in comparison to the end-bulk non-diagonal interaction, the bulk-bulk non-diagonal interaction has only minor impact on spectrum and transport properties of narrow ACNs.


For the non-diagonal end-bulk interaction, we find, as explicitly evaluated in \App{sec:CalcVeb},
\begin{eqnarray}\nonumber&\left\langle \vec{N}\ \vec{\sigma}^{{ \edge}}=(\sigma_+,\sigma_-)\ \vec{m}\left|\hV^{{\eb}}_{n\rho\rho}\right|\vec{N}'\ \vec{\sigma}^{{ \edge}}{}'=(\sigma'_+,\sigma'_-)\ \vec{m}'\right\rangle
\\\nonumber&=t\sum_{p}\delta_{\vec{N},\vec{N'}+\vec{e}_{\sigma'_p}-\vec{e}_{\sigma_p}}\delta_{\vec{m}_c,\vec{m}'_c}\delta_{\sigma_p,-\sigma'_{p}}\delta_{\sigma_{\bar{p}},\sigma'_{\bar{p}}}\,\left[\delta_{p,-}(-1)^{N'_\uparrow}-\delta_{p,+}(-1)^{N'_\downarrow}\right]\\\ &\times\mathrm{sgn}(\sigma_p)
\prod_qF\left(p\,\mathrm{sgn}(\sigma'_p)\sqrt{2/n_q},m_{sq},m'_{sq}\right).\label{V_e-b-final}\end{eqnarray}
The action of this scattering process is to flip the localized spin at the $p$ end, $\sigma'_p\to\sigma_p\overset{!}{=}-\sigma'_p$, while at the same time a bulk spin must be inverted to preserve the spin-projection $S_z$: $N'_{\sigma'_p}\to N_{\sigma'_p} \overset{!}{=}N'_{\sigma'_p}+1\,,\ N'_{-\sigma'_p}\to N_{-\sigma'_p}\overset{!}{=}N'_{-\sigma'_p}-1$. The localized spin at the other end, i.e., the $\bar{p}\equiv-p$ end, must be conserved: $\sigma_{\bar{p}}=\sigma'_{\bar{p}}$. As a result, the degeneracy between the lowest lying states of $\hat H_0+\hat V_{\rho\rho} $ is removed; moreover, the spin-charge separation gets smeared \cite{Koller09}.\\
\section{Minimal model for the lowest lying states\label{sec:minmod}}
The low energy spectrum resulting from the numerical diagonalization of $\hat H_{\odot}$, Eq. (\ref{ACN-ham}),  is discussed in Ref. \cite{Koller09}. Here to explain
 how the lowest-lying states in the truncated eigenbasis Eq. (\ref{eq:eigenstates_H0Vrr}) transform under the influence of $\hV_{n\rho\rho}$ it is sufficient to restrict to a minimal set of states: 
For the even fillings, \mbox{$N_c=2n,\,n\in\mathbb{N}$}, we have to take into account twelve states, allowing up to one fermionic excitation. The reason is that without an unpaired bulk spin no mixing can take place\footnote{For transport, this is fatal, as introducing the end spins degree of freedom a priori yields four identical, completely decoupled channels. This makes the kinetic equations ill-defined. Only the end-bulk interaction induced mixing guarantees a unique solution for the transport problem.}. For the odd fillings, $N_c=2n+1$, it is enough to include the eightfold degenerate ground state of $\hH_0+\hV_{\rho\rho}$. To preserve lucidity, bosonically excited states are left out from our analysis, as they do not change qualitatively the mechanisms behind the observed effects. The $\vec{N}$ and $\vec{m}_c$ conserving bulk-bulk scattering $\hV^{{\bb}}_{n\rho\rho}$ can also be disregarded: besides for slight shifts in energy, forming linear combinations of states differing just in $\vec{m}_s$ but identical in $\vec{N}$, $\vec{m}_c$ has not much impact. The restriction to the minimal model applies only for the subsequent analytics. Concerning the numerical calculations shown in Sec. \ref{sec:transport}, an energy cutoff of $1.9\epsilon_0$ above the ground states was used, including every energetically allowed bosonic or fermionic excitation.

\subsection{Even electron fillings\label{evenN}}

For even electron fillings, $N_c=2n$, we want to restrict to the following subset of the states described by Eq. (\ref{eq:eigenstates_H0Vrr}):
\begin{eqnarray*}\vec{N}=(N_\uparrow,N_\downarrow)&\in&\{(n,n),(n\pm1,n\mp1)\}\quad n\in\mathbb{N},\\\vec{\sigma}^{{ \edge}}=({\sigma}^{{ \edge}}_+,{\sigma}^{{ \edge}}_-)&\in&\{(\uparrow,\uparrow),(\uparrow,\downarrow),(\downarrow,\uparrow),(\downarrow,\downarrow)\},\\
\vec{m}=(\vec{m}_c,\vec{m}_s)&=&(\vec{0},\vec{0}).\end{eqnarray*}
To abbreviate our notation we introduce \begin{equation*}\label{statedef}\left|\vec{N},\vec{\sigma}^{{ \edge}},\vec{0}\right\rangle:=\left\lbrace\begin{array}{ll}\left|{\sigma}^{{ \edge}}_-,\,\uparrow\downarrow\,,{\sigma}^{{ \edge}}_+\right\rangle & N_\uparrow=N_\downarrow=n,\\\left|{\sigma}^{{ \edge}}_-,\,\uparrow\uparrow\,,{\sigma}^{{ \edge}}_+\right\rangle & N_\uparrow=n+1,\ N_\downarrow=n-1,\\\left|{\sigma}^{{ \edge}}_-,\,\downarrow\downarrow\,,{\sigma}^{{ \edge}}_+\right\rangle & N_\uparrow=n-1,\ N_\downarrow=n+1.\end{array}\right.\end{equation*}
Notice that the second and the third case describes fermionically excited states.
Building now all possible combinations for our minimal set of states for even fillings, we get the following set of possibilities:

$\bullet\,$ \textit{$S_z=0$: \bf{four states $|a\rangle,\,|b\rangle,\,|c_{\pm}\rangle$}}\begin{eqnarray*}
\left|\uparrow,\uparrow\downarrow,\downarrow\right>=:|a\rangle,\ \,&\!\!&\quad\left|\downarrow,\uparrow\downarrow,\uparrow\right>=:|b\rangle,\\\left|\uparrow,\downarrow\downarrow,\uparrow\right>=:|c_{+}\rangle,&\!\!&\quad\left|\downarrow,\uparrow\uparrow,\downarrow\right>=:|c_{-}\rangle,\end{eqnarray*}

$\bullet\,$ \textit{$S_z=\pm\hbar$: \bf{six states $|d_{\sigma\sigma}\rangle,\,|f_{\sigma\sigma}\rangle,\,|g_{\sigma\sigma}\rangle\ (\sigma\in\{\uparrow,\downarrow\})$}}\begin{eqnarray*}
\left|\uparrow,\uparrow\downarrow,\uparrow\right>=:|d_{\uparrow\uparrow}\rangle,&\!\!&\quad\left|\downarrow,\uparrow\downarrow,\downarrow\right>=:|d_{\downarrow\downarrow}\rangle,\\\left|\uparrow,\uparrow\uparrow,\downarrow\right>=:|f_{\uparrow\uparrow}\rangle,&\!\!&\quad\left|\uparrow,\downarrow\downarrow,\downarrow\right>=:|f_{\downarrow\downarrow}\rangle,\\\left|\downarrow,\uparrow\uparrow,\uparrow\right>=:|g_{\uparrow\uparrow}\rangle,&\!\!&\quad\left|\downarrow,\downarrow\downarrow,\uparrow\right>=:|g_{\downarrow\downarrow}\rangle,\end{eqnarray*}

\begin{figure}
\begin{center}\includegraphics[width=0.98\columnwidth]{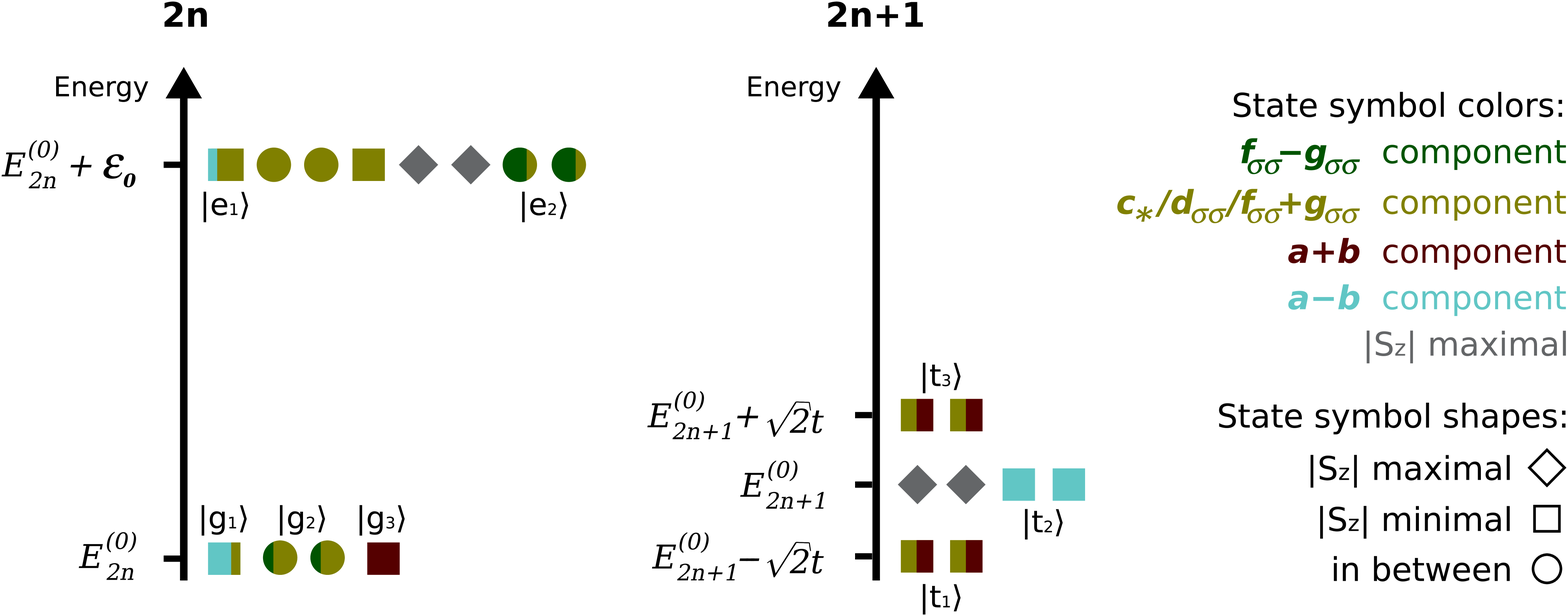}\end{center}
\caption{\label{levels}Energy landscape for a minimal set of lowest ACN eigenstates for even ($N_c=2n$) and odd ($N_c=2n+1$) electron fillings, in accordance with \Tabs{states-2n}{states-2n+1}. To visualize the relevant contributions in the composition of the eigenstates, different colored and shaped symbols were used. The states relevant for later considerations are labelled.\bigskip}
\end{figure}

$\bullet\,$ \textit{$S_z=\pm2\hbar$: \bf{two states $|\sigma,\sigma\sigma,\sigma\rangle\ (\sigma\in\{\uparrow,\downarrow\})$}}\begin{equation*}
\left|\uparrow,\uparrow\uparrow,\uparrow\right>,\quad\left|\downarrow,\downarrow\downarrow,\downarrow\right>.\end{equation*}
\noindent There are four degenerate ground states $|a\rangle,\,|b\rangle,\,|d_{\sigma\sigma}\rangle$ with energy $E^{(0)}_{N_c}=E^{(0)}_{2n}$, while the remaining singly fermionic excited states $|f_{\sigma\sigma}\rangle,\,|g_{\sigma\sigma}\rangle,\,|\sigma,\sigma\sigma,\sigma\rangle$ have an energy $E^{(f)}_{2n}$.\\
The end-bulk interaction can only mix states with same spin-projection $S_z$. To the highest values, $S_z=\pm2\hbar$, it belongs only one state and thus no mixing occurs. In contrast, the three states $|d_{\sigma\sigma}\rangle,\,|f_{\sigma\sigma}\rangle,\,|g_{\sigma\sigma}\rangle$ with $S_z=\sgn(\sigma)\hbar$ get coupled to each other. Also the four states $|a\rangle,\,|b\rangle,\,|c_{\pm}\rangle$ with $S_z=0$ can in principle transform into one another under the influence of end-bulk scattering. With help of Eq. (\ref{V_e-b-final}) we can set up the corresponding blocks of the full Hamiltonian $\hH_\odot$ in the truncated eigenbasis (assuming $n$ even\footnote{For odd $n$ the sign of the off-diagonal entries must be inverted.}):
\begin{table}
\begin{center}
\begin{tabular}{|r@{\quad:\quad}ll@{\quad}cc|}\hline
\multicolumn{5}{|c|}{$N_c=2n$}\\
\textsc{energy} & \textsc{eigenstate\ (not\ normalized)} &\textsc{abbr.}&&\textsc{spin} $S_z\ [\hbar]$\\\hline\hline
$\xi_{+-}(4t)\approx 0$ & $\Biggl.\frac{2\sqrt{2}t}{\xi_{--}(4t)}(|a\rangle-|b\rangle)-(|c_+\rangle+|c_{-}\rangle)\Biggr.$&\geins&\includegraphics[width=4mm]{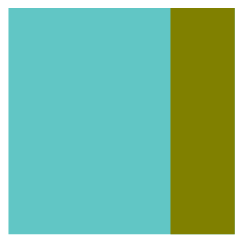}&$0$\\
$\xi_{+-}(2\sqrt{2}t)\approx 0$& $\Biggl.\frac{2\sqrt{2}t}{\xi_{--}(2\sqrt{2}t)}|d_{\sigma\sigma}\rangle+(|f_{\sigma\sigma}\rangle-|g_{\sigma\sigma}\rangle)\Biggr.$&\gzwei&\includegraphics[width=4.2mm]{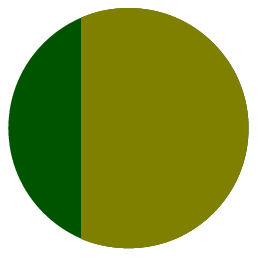}&${\sgn(\sigma)}$\\
$0$ & $\Biggl.\sqrt{\frac{1}{2}}(|a\rangle+|b\rangle)\Biggr.$&\gdrei&\includegraphics[width=4mm]{q_rot.eps}&$0$
\\\hline
$\xi_{++}(4t)\approx\varepsilon_0$ & $\Biggl.\frac{2\sqrt{2}t}{\xi_{-+}(4t)}\left(|a\rangle-|b\rangle\right)-\left(|c_+\rangle+|c_{-}\rangle\right)\Biggr.$&\eeins{}&\includegraphics[width=4mm]{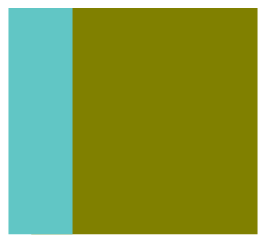}&$0$\\
$E^{(f)}_{2n}\approx\varepsilon_0$ & $\Biggl.\frac{1}{\sqrt{2}}\left(|f_{\sigma\sigma}\rangle+|g_{\sigma\sigma}\rangle\right)\Biggr.$&&\includegraphics[width=4.2mm]{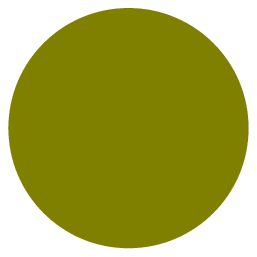}&${\sgn(\sigma)}$\\
$E^{(f)}_{2n}\approx\varepsilon_0$ & $\Biggl.\sqrt{\frac{1}{2}}(|c_+\rangle+|c_-\rangle)\Biggr.$&&\includegraphics[width=4mm]{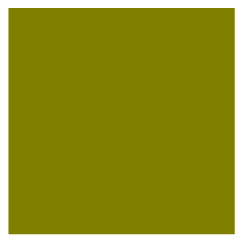}&$0$\\
$E^{(f)}_{2n}\approx\varepsilon_0$ & $\Biggl.\left|\,\sigma,\sigma\sigma,\sigma\right\rangle\Biggr.$&&\includegraphics[width=4.4mm]{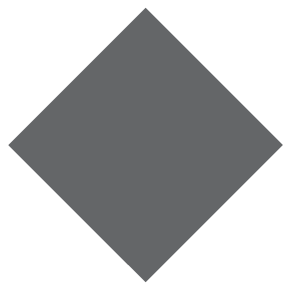}&$2{\sgn(\sigma)}$\\
$\xi_{++}(2\sqrt{2}t)\approx\varepsilon_0$ & $\Biggl.\frac{2\sqrt{2}t}{\xi_{-+}(2\sqrt{2}t)}|d_{\sigma\sigma}\rangle+(|f_{\sigma\sigma}\rangle-|g_{\sigma\sigma}\rangle)\Biggr.$&\ezwei&\includegraphics[width=4.2mm]{c_gruenoliv.eps}&${\sgn(\sigma)}$\\
\hline
\end{tabular}\vskip 0.4cm
\caption{The minimal model consists of twelve different states for even electron fillings of the ACN ($\sigma\in\{\uparrow,\downarrow\}$). We have set $E^{(0)}_{2n}=0$. Importantly, fermionically excited states mix with non excited ones. States relevant for explanations in the main text are marked with text labels. The color and shape of the symbols encodes the nature of the state, classified by spin-projection and behavior under exchange of the end spins. The boxes indicate $S_z=0$, related colors are red (symmetric component $|a\rangle+|b\rangle$) and blue (antisymmetric combination $|a\rangle-|b\rangle$). For the disks $|S_z|=\hbar$, colors olive (symmetric components) and green (antisymmetric: $|f_{\sigma\sigma}\rangle-|g_{\sigma\sigma}\rangle$). Grey diamonds: $|S_z|=2\hbar$.\label{states-2n}}
\end{center}\end{table}
\begin{eqnarray*}&&\quad\ \ |a\rangle\quad\ \,|b\rangle\quad\ |c_+\rangle\,\quad|c_-\rangle\\
\left(\hH_\odot\right)_{N_c=2n,S_z=0}&=&\left(\begin{array}{cccc}E^{(0)}_{2n}&0&-t&-t\\0&E^{(0)}_{2n}&+t&+t\\-t&+t&E^{(f)}_{2n}&0\\-t&+t&0&E^{(f)}_{2n}\end{array}\right)\begin{array}{l}|a\rangle\\|b\rangle\\|c_+\rangle\\|c_-\rangle\ ,\end{array}\\&&\\&&\ \quad\ |d_{\sigma\sigma}\rangle\ \,|f_{\sigma\sigma}\rangle\ \,|g_{\sigma\sigma}\rangle\\
 \left(\hH_\odot\right)_{N_c=2n,S_z=\pm\hbar}&=&\,\left(\begin{array}{ccc}E^{(0)}_{2n}&+t&-t\\ +t&E^{(f)}_{2n}&0\\-t&0&E^{(f)}_{2n}\end{array}\right)\begin{array}{l}|d_{\sigma\sigma}\rangle\\|f_{\sigma\sigma}\rangle\\|g_{\sigma\sigma}\rangle\ .\end{array}\end{eqnarray*}
Diagonalization leads to the eigenstates and eigenenergies listed in \Tab{states-2n}. We employ there the abbreviation\numparts\begin{equation}\label{xi}\xi_{\alpha\alpha'}(\gamma)=\frac{1}{2}\left(E^{(f)}_{2n}+\alpha E^{(0)}_{2n}+\alpha'\sqrt{\left(E^{(f)}_{2n}-E^{(0)}_{2n}\right)^2+\gamma^2}\right),\end{equation} with $\alpha,\alpha'\in\{\pm1\}$. Obviously, $\xi_{++}(\gamma)>\xi_{+-}(\gamma)\,,\quad\xi_{-+}(\gamma)>\xi_{--}(\gamma)$, and as in our context $\gamma\simeq t$, and hence $\gamma\ll E^{(f)}_{2n}-E^{(0)}_{2n}$ holds, we can rely on the relations \begin{equation}\xi_{++}(\gamma)\approx E^{(f)}_{2n}\,,\quad\xi_{+-}(\gamma)\approx E^{(0)}_{2n}\,,\quad\xi_{-+}(\gamma)\gg t\,,\quad\xi_{--}(\gamma)\ll t.\label{smallbig}\end{equation}\endnumparts
The resulting energy landscape is sketched on the left side of \Fig{levels}, where we used differently colored and shaped symbols to indicate the composition of states.
In our simple model, we find then from \Tab{states-2n} that the interaction has hardly lifted the degeneracies between the various states.
It can be verified with \Eq{xi} that there is a slight splitting in the ground states, such that their energy grows from \geins{} to \gdrei{}.
Also the excited states, of which only the two labelled ones turn out to be important for later explanations, are listed increasing in energy. Crucial is the mixing of states with different bulk and end configurations. We can single out linear combinations which are either symmetric or antisymmetric under exchange of the end spins.
For the features we will observe in transport, the decisive entanglement is the one between the four $S_z=0$ states $|a\rangle,\,|b\rangle,\,|c_+\rangle,\,|c_-\rangle$,
leading to two states containing the antisymmetric combination $|a\rangle-|b\rangle$: A ground state \geins{} with small contribution of the fermionically excited states $|c_+\rangle+|c_-\rangle$,
and an excited state \eeins{} where those dominate [as found with \Eq{smallbig}].

\subsection{Odd electron fillings\label{oddN}}
Here, due to the fact that with $N_c=2n+1$ we necessarily always have an unpaired spin, it is sufficient to consider merely the ground states with energy $E^{(0)}_{N_c}=E^{(0)}_{2n+1}$ emerging from Eq. (\ref{eq:eigenstates_H0Vrr}):
\begin{eqnarray*}\vec{N}=(N_\uparrow,N_\downarrow)&\in&\{(n+1,n),(n,n+1)\}\quad n\in\mathbb{N},\\\vec{\sigma}^{{ \edge}}=({\sigma}^{{ \edge}}_+,{\sigma}^{{ \edge}}_-)&\in&\{(\uparrow,\uparrow),(\uparrow,\downarrow),(\downarrow,\uparrow),(\downarrow,\downarrow)\},\\
\vec{m}=(\vec{m}_c,\vec{m}_s)&=&(\vec{0},\vec{0}).\end{eqnarray*}
Again we abbreviate our notation and introduce\begin{equation*}\label{statedef_odd}\left|\vec{N},\vec{\sigma}^{{ \edge}},\vec{0}\right\rangle:=\left\lbrace\begin{array}{ll}\left|{\sigma}^{{ \edge}}_-,\,\uparrow\,,{\sigma}^{{ \edge}}_+\right\rangle & N_\uparrow=n+1,\ N_\downarrow=n,\\\left|{\sigma}^{{ \edge}}_-,\,\downarrow\,,{\sigma}^{{ \edge}}_+\right\rangle & N_\uparrow=n,\ N_\downarrow=n+1.\end{array}\right.\end{equation*}
We get the following set of possibilities:

$\bullet\,$ \textit{$S_z=\pm\hbar/2$: \bf{six states $|a_\sigma\rangle,\,|b_\sigma\rangle,\,|c_\sigma\rangle\ (\sigma=\uparrow\leftrightarrow\bar{\sigma}=\downarrow,\,\sigma=\downarrow\leftrightarrow\bar{\sigma}=\uparrow), $}}\[
\left|\uparrow,\sigma,\downarrow\right>=:|a_\sigma\rangle,\quad\left|\downarrow,\sigma,\uparrow\right>=:|b_\sigma\rangle,\quad\left|\sigma,\bar{\sigma},\sigma\right>=:|c_\sigma\rangle,\]

$\bullet\,$ \textit{$S_z=\pm3\hbar/2$: \bf{two states $|\sigma,\sigma,\sigma\rangle\ (\sigma\in\{\uparrow,\downarrow\})$}}\begin{equation*}
\left|\uparrow,\uparrow,\uparrow\right>,\quad\left|\downarrow,\downarrow,\downarrow\right>.\end{equation*}
In the case of $S_z=\pm3\hbar/2$ there is only one state each.\\For the three states with $S_z=\pm\hbar/2$, from  Eqs. (\ref{eq:H0Vrr_diag}) and (\ref{V_e-b-final}) the following mixing matrix is found (still $n$ is assumed even):\newpage
\begin{eqnarray*}&\quad\quad|a_\sigma\rangle\quad\quad|b_\sigma\rangle&\quad\ |c_\sigma\rangle\\
\left(\hH_\odot\right)_{N_c=2n+1,S_z=\pm\hbar/2}=&\left(\begin{array}{c}E^{(0)}_{2n+1}\\0\\-t\end{array}\right.\begin{array}{c}0\\E^{(0)}_{2n+1}\\-t\end{array}&\left.\begin{array}{c}-t\\-t\\E^{(0)}_{2n+1}\end{array}\right)\begin{array}{l}|a_\sigma\rangle\\|b_\sigma\rangle\\|c_\sigma\rangle\ .\end{array}\\
\end{eqnarray*}
\begin{table}
\begin{center}
\begin{tabular}{|r@{\quad:\quad}ll@{\qquad}cc|}\hline
\multicolumn{5}{|c|}{$N_c=2n+1$}\\
\textsc{energy} & \textsc{eigenstate\ (normalized)} &\textsc{abbr.}&&\textsc{spin} $S_z\ [\hbar]$\\\hline\hline
$-\sqrt{2}t$ & $\Biggl.\frac{1}{2}(|a_\sigma\rangle+|b_\sigma\rangle)+\sqrt{\frac{1}{2}}|c_\sigma\rangle\Biggr.$&\teins${}$&\includegraphics[width=4mm]{q_olivrot.eps}&${\sgn(\sigma)}/2$\\
$0$ & $\sqrt{\frac{1}{2}}(|a_\sigma\rangle-|b_\sigma\rangle)\Biggr.$&\tzwei${}$&$\includegraphics[width=4mm]{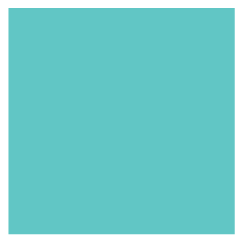}$&${\sgn(\sigma)}/2$\\
$0$ & $\Biggl.\left|\,\sigma,\sigma,\sigma\right\rangle\Biggr.$&&\includegraphics[width=4.4mm]{d_grau.eps}&$3{\sgn(\sigma)}/2$\\
$+\sqrt{2}t$ & $\Biggl.\frac{1}{2}(|a_\sigma\rangle+|b_\sigma\rangle)-\sqrt{\frac{1}{2}}|c_\sigma\rangle\Biggr.$&\tdrei${}$&\includegraphics[width=4mm]{q_olivrot.eps}&${\sgn(\sigma)}/2$\\\hline
\end{tabular}\vskip 0.4cm
\caption{Lowest lying eigenstates of an ACN filled with an odd number of electrons ($\sigma\in\{\uparrow,\downarrow\}$). Due to spin-degeneracy, the total number of possible states is eight. The two states with $|S_z|=3\hbar/2$ are marked by grey diamonds. Red and olive boxes stand for the symmetric components $|a_\sigma\rangle+|b_\sigma\rangle$ and $|c_\sigma\rangle$, respectively. Blue boxes label the antisymmetric combinations $|a_\sigma\rangle-|b_\sigma\rangle$. Notice that all states behave purely symmetric or antisymmetric with respect to end spin exchange.\label{states-2n+1}}
\end{center}
\end{table}
The matrix is easily diagonalizable and yields eigenstates according to \Tab{states-2n+1} at three distinct eigenenergies (compare also to \Fig{levels}, right).
Notice that for the odd filling all emerging states are purely symmetric or antisymmetric under exchange of the two end spins. Thereby, the symmetric states \teins{} and \tdrei{} essentially have the same tunneling properties, because they only differ by the sign in front of $|c_\sigma\rangle$. It is of crucial importance that the  state \tzwei{} is formed by the \emph{antisymmetric} combination $|a_\sigma\rangle-|b_\sigma\rangle$. Comparing the definition of the $2n$ states $|a\rangle,\,|b\rangle$ and $|a_\sigma\rangle,\,|b_\sigma\rangle$, we see that from the $2n$ ground state \gdrei{}, a tunneling event can \textit{exclusively} lead to one of the $2n+1$ states \teins{} or \tdrei{}. Via their $|c_\sigma\rangle$ components, these connect to $|c_+\rangle+|c_-\rangle$ as well as to $|d_{\sigma\sigma}\rangle$, and thus to all the other labelled $2n$ states from \Tab{states-2n}, but the link to \geins, \ezwei{} is weak due to \Eq{smallbig}. This will be the key ingredient for the explanation of the stability diagrams in \Sct{sec:transport}.

\section{Spin-dependent transport across quantum-dot ACNs\label{sec:transport}}

Looking solely to the spectrum, there is no demand for distinguishing between symmetry or antisymmetry of a certain state under the exchange of end spins. In transport, however, this property turns out to lead to tremendous effects.
The case of an unpolarized set-up was discussed in Ref. \cite{Koller09}, where it was found that end-bulk entanglement leads to pronounced negative differential conductance (NDC) lines occurring for a completely symmetric setup. In this work the focus is on spin-dependent transport for collinear lead magnetizations. Strikingly, all the NDC features observed for the unpolarized set-up \emph{vanish} for anti-parallel contact magnetization while they persist for the parallel case. As a consequence we predict negative tunneling magneto-resistance (TMR) within a narrow region along the edges of the Coulomb diamonds for even fillings.\\

Unless specified differently, we have employed the following parameters for all viewed plots:\begin{center}
\begin{tabular}{rll}
Energy cutoff &$E_{{ max}}$&$1.9\varepsilon_0$,\\
Thermal energy&$k_BT$& $0.01\,$meV,\\
Charging energy\footnotemark & $W_0$ & $2.31$\,meV,\\
Ribbon length &$L_y$& $572$\,nm,\\
Level spacing & $\epsilon_0$ & $2.93\,$meV,\\
Ribbon width &$L_x$& $7.8\,$nm,\\
Bulk-bulk exchange&$u$&$0.036$\,meV,\\
End-bulk exchange&$t$&$0.21$\,meV. 
\\
Polarization strength&$P$&$0.8$\\&&
\end{tabular}\footnotetext{A dielectric constant $\epsilon=1.4$ was assumed.}\end{center}
The values of charging energy, bulk-bulk and end-bulk exchange-coupling were numerically verified for ribbon widths ranging from $5-20\,$nm.\\ 

Throughout this work we assume that the coupling between the electronic reservoirs, i.e. the contacts, and the ACN is weak. Under such condition, the total system
is described by
\begin{equation*}
\hH=\hH_\odot+\hH_{\leads}+\hH_T -e\alpha V_{\gate}\hat{\mathcal{N}}_c,
\end{equation*}
with the ACN-Hamiltonian $\hH_\odot$ given in Eq. (\ref{ACN-ham}).
Further, the contacts are described by $\hH_{\leads}=\sum_{lq}\sum_{\sigma}(\epsilon_q-\mu_{l})\hc^\dag_{l\sigma q}\hc_{l\sigma q}$, with $\hc_{l\sigma q}$ annihilating an electron in lead $l$ of kinetic energy $\epsilon_q$. The chemical potential $\mu_l$ differs for the left and right contact by $eV_\bias$, with $V_\bias$ the applied bias voltage.
Next, $H_T=\sum_{l\sigma}\int\!\! d^3 r\left(T_l(\vec{r})\hpsi^\dag_{\sigma}(\vec{r})\hphi_{l\sigma}(\vec{r})+h.c.\right)$ describes tunneling between ACN and contacts, where $T_l(\vec{r})$ is the in general position dependent tunneling coupling and $\hpsi_{\sigma}(\vec{r})$ the ACN bulk electron operator as given in Eq. (\ref{eq:3Dto1D}). The lead electron operator is $\hphi_{l\sigma_l}(\vec{r})=\sum_{q}\varphi_{lq}(\vec{r})\hc_{l\sigma_lq}$ with $\varphi_{lq}(\vec{r})$ denoting the wave function of the contacts. Finally, the potential term describes the influence
 of a capacitively applied gate voltage $(0\le \alpha \le 1)$.
Due to the condition that the coupling between ACN and the contacts is weak,
we can calculate the stationary current by solving a master equation for the
reduced density matrix to second order in the tunneling coupling. As this is a standard procedure, we
refer to previous works~\cite{Mayrhofer06,Koller07,Hornberger08} for details about the method.

Besides the energy spectrum, the other system specific input required for transport calculations are the tunneling matrix elements of the ACN bulk electron operator,
\begin{eqnarray*}&\left\langle\vec{N}\vec{\sigma}^{\small \edge}\vec{m}\left|\hPsi_{\sigma}(\vec{r})\right|\vec{N}'\vec{\sigma}^{\small \edge}{}'\vec{m}'\right\rangle=\frac{1}{2}\delta_{\vec{N},\vec{N}'+\vec{e}_{\sigma}}\delta_{\vec{\sigma}^{\small \edge},\vec{\sigma}^{\small \edge}{}'}\left(-1\right)^{\delta_{\sigma,\downarrow}N_\uparrow}\\&\times\sum_{Fpr}f_{Fpr}\varphi_{Fp}(\vec{r}) e^{i \frac{\pi}{L_y}({N}_{\sigma }+\frac{1}{2})ry}\prod_{q>0}\prod_{j=s,c}F(\lambda^{jq}_{r\sigma}(y),m_{jq},m'_{jq}),\end{eqnarray*} with the function $F(\lambda,m,m')$ and the parameter $\lambda^{jq}_{r\sigma}(y)$ as given in \App{app:matrixelements-bulk} in Eq. (\ref{eq:Fvonlambda1}) respectively Eq. (\ref{eq:lambda_einfach}). We omit here the calculation of this identity, because the fermionic contribution follows straightforwardly from Eqs. (\ref{eq:bosident})-(\ref{eq:K_rsF}), while the bosonic contribution emerges in the same manner as for the more complicated matrix elements involving more than one electron operator evaluated in \App{app:matrixelements-bulk}. Moreover, the detailed derivation of the corresponding expression for SWCNTs can be found in Ref.~\cite{Mayrhofer07}.

\subsection{Collinearly spin-polarized transport without magnetic field\label{sec:NDC}}
\begin{figure}
\begin{center}
\includegraphics[width=12.6cm]{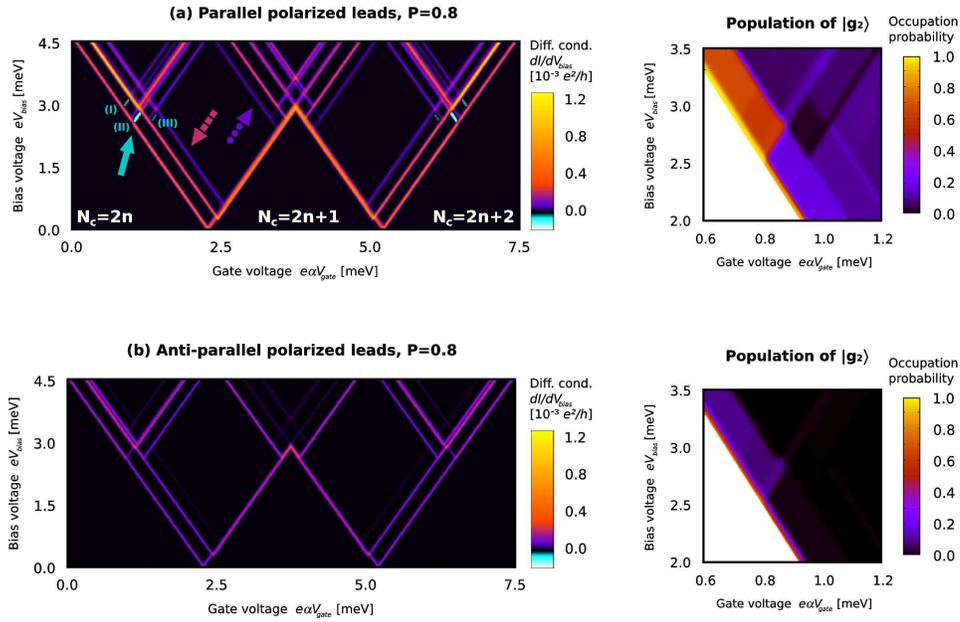}\vskip 0.2cm
\caption{(a) Left: Differential conductance for an ACN transport setup with parallel lead polarizations. Strong negative differential conductance (NDC) arises if the external voltages are adjusted such that a transition from \teins{} to \eeins{} is allowed, while a transition from \geins{} to \tzwei{} is forbidden. (a) Right: Occupation probability of the trapping state \geins{} around the region exhibiting various NDC features. Notice that no numerically stable data can be obtained inside the Coulomb diamond.\label{ACN-80}
(b) Left: Differential conductance for an ACN quantum-dot connected to anti-parallel polarized leads. The number of visible transition lines is strongly decreased as compared to the parallel case \Fig{ACN-80}a (left). Further, \emph{all NDC features have vanished}.
\label{ACN-AP}
(b) Right: Occupation probability of the state \geins{} for the same bias and gate range as in \Fig{ACN-80}a (right). For an anti-parallel contact configuration, the population of the state is strongly decreased.
}
\end{center}\end{figure}

In \Fig{ACN-80}, left, (a) and (b) we show the stability diagrams obtained for an ACN coupled to leads polarized in parallel and in anti-parallel, respectively.
Due to electron-hole symmetry around the $2n$ filling, the transport characteristics are mirror-symmetric with respect to the central diamond. As anticipated at the beginning of this section, 
end states leave various signatures in the parallel case, \Fig{ACN-80}, left, (a) which are absent in the anti-parallel configuration, \Fig{ACN-80}, left, (b).
In the following we give an  explanation for all effects indicated in \Fig{ACN-80}, left, (a)
 based on the minimal set of states discussed in \Sct{evenN}, \Tabs{states-2n}{states-2n+1}. Let us shortly recall their properties and alongside explain in which points we have to expect discrepancy with respect to the full set of real eigenstates which was employed for all calculations:\\
Firstly, the eight states from \Tab{states-2n+1}, \teins, \tzwei, \tdrei{} and $|\sigma,\sigma,\sigma\rangle$, will in the following frequently be called the \emph{lowest lying $2n+1$ states}. They occur at only three distinct energies $\pm\sqrt{2}t,\ 0.0$. Inclusion of excitations within an energy cutoff of $1.9\,\varepsilon_0$ slightly lifts the degeneracy of \tzwei{} and $|\sigma,\sigma,\sigma\rangle$, and introduces eight high lying excited states which are almost degenerate. Secondly, for what concerns the even fillings, we refer to \geins, \gzwei, \gdrei{} as \emph{2n ground states}. The fact that they are almost, but not perfectly degenerate, and that their energy grows from \geins{} to \gdrei{}, is not changed upon the inclusion of further excitations and plays some role in the following. Moreover, mixing between the states from \Tab{states-2n} and bosonically excited $2n$ states takes place in general, but actually preserves the types of linear combinations occurring in \Tab{states-2n}, which is the relevant point for our explanations. In summary, the main effect of inclusion of excitations within an energy cutoff of $1.9\,\varepsilon_0$ is the lifting of the degeneracies among the \emph{excited} $2n$ states. In fact, the lowest lying excited state will be of the same nature as \eeins{}, and the separation to the state corresponding to \ezwei{} exceeds $\sqrt{2}t$, thus it is well resolvable. With this additional information to \Tabs{states-2n}{states-2n+1} we can now start to explain the features marked by the different arrows in \Fig{ACN-80} (left).\\

\includegraphics[width=0.8cm]{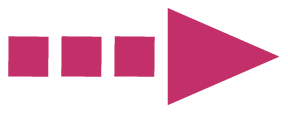}\ The dashed red arrow points towards a triple of three parallel lines which are split by $\sqrt{2}t$. Those mark transitions from the $2n$ ground states to the $2n+1$ lowest lying states. Hereby, the antisymmetric state \tzwei{}, associated to the second line of the triple, is special, because it is the only one strongly connected to the $2n$ state \geins. 
The first line of the triple is the \gdrei$\to$\teins{} ground state transition line.\\

\includegraphics[width=0.8cm]{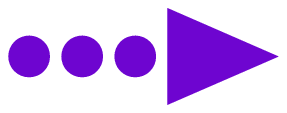}\ The blue dotted arrow marks the lines around the tip of the Coulomb diamond. Those appear in four clearly distinct positions, separated by about $\sqrt{2}t$ . The lower lying triple of lines arises from transitions of the lowest lying $2n+1$ states to the $2n+2$ ground states. By coincidence of parameters, the highest line, which is split, follows also in a distance of about $\sqrt{2}t$ and marks transitions from $2n$ ground states to the aforementioned higher lying excited $2n+1$ states arising upon inclusion of bosonic excitations.\\

\begin{figure}\begin{center}
\includegraphics[width=0.99\columnwidth]{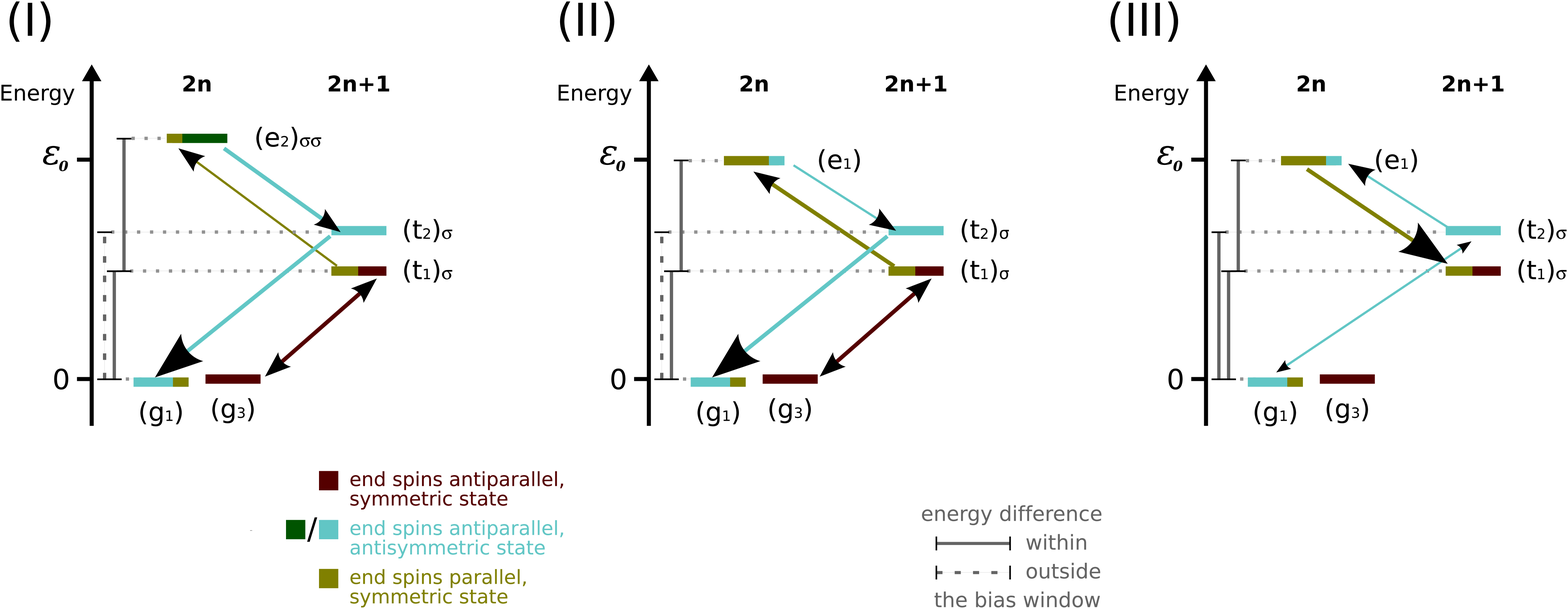}{\ \ }\vskip 0.5cm
\caption{Schematic explaining the mechanisms causing the NDC features (I), (II) and (III) in \Fig{ACN-80} (left). Only states and transitions relevant for the NDCs are drawn. The crucial transition is marked by a big arrow head. (I)/(II) Opening of the channel \teins$\to$\ezwei{}, respectively \teins$\to$\eeins{}, leads to a decay into the trapping state \geins, depleting the transport channel \gdrei$\leftrightarrow$\teins. (III) Opening of the channel \tzwei$\to$\eeins{} depletes the transport channel \geins$\leftrightarrow$\tzwei.\label{scheme}}\vspace{-0.4cm}\end{center}
\end{figure}
\includegraphics[width=0.8cm]{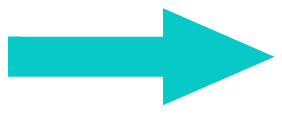}\ The solid green arrow highlights the negative differential conductance features (I), (II) and (III). The former two originate from trapping in the state \geins, while the latter occurs due to depletion of the transport channel \tzwei$\leftrightarrow$\geins.\\The mechanisms work as follows:\\
The NDCs (I) and (II) mark the opening of the $2n+1\to 2n$ back-transition channels \teins$\to$\ezwei{} and \teins$\to$\eeins{}, respectively. The situation is sketched in \Fig{scheme}. Once they get populated, from both of these excited $2n$ states the system can decay into any of the lowest lying $2n+1$ states, and in particular there is a chance to populate the antisymmetric state \tzwei{}. This state is strongly connected to the $2n$ ground state \geins{}, which contains a large contribution of the antisymmetric combination $|a\rangle-|b\rangle$. But in the region where the NDCs occurs, the forward channel \geins$\to$\tzwei{} is not yet within the bias window such that \geins{} serves as a trapping state. \Fig{ACN-80}a (right) confirms this explanation: the population of the state \geins{} is strongly enhanced in the concerned region where the back-transitions \teins$\to$\ezwei{} and \teins$\to$\eeins{} can take place, while the forward transition \geins$\to$\tzwei{} is still forbidden.\\
NDC (III) belongs to the back-transition \tzwei$\to$\eeins{}, which is a weak channel because \tzwei{} is a purely antisymmetric state, while the antisymmetric contribution in \eeins{} is rather small. From time to time, nevertheless the transition will take place, and once it happens the system is unlikely to fall back to \tzwei, but will rather change to a symmetric $2n+1$ state. 
Thus the state \tzwei{} is depleted, and with it the transport channel \tzwei$\leftrightarrow$\geins, which leads to NDC. The statement can also be verified from the plot of the occupation probability for \geins, \Fig{ACN-80} (right): a pronounced dark region of decreased population follows upon the NDC transition.\\ 


Major changes in the stability diagram of the ACN are observed for anti-parallel contact configuration, \Fig{ACN-AP}b (left): Compared to \Fig{ACN-80}a (left), various transitions
lines are suppressed and the NDC features have vanished. The reason is that an anti-parallel contact configuration as drawn in \Fig{graphene} favors  (g$_2)_{\downarrow\downarrow}$ as $2n$ ground state, because in-tunneling of $\downarrow$ - electrons and subsequent out-tunneling of $\uparrow$ - electrons is preferred. As a consequence, all transport channels related to \geins{} and \gdrei{} are of minor relevance, which weakens various transition lines and in particular destroys the NDCs effects:  in the anti-parallel configuration, the occupation of the trapping state \geins{} is significantly lowered, as seen in \Fig{ACN-AP}b (right).\\
In detail, starting from the $2n$ ground state \gdrei$=\frac{1}{\sqrt{2}}(|a\rangle+|b\rangle)$, in-tunneling of a majority ($\downarrow$ -) electron from the source takes the system to (t$_1)_\downarrow=\frac{1}{\sqrt{2}}(|a_\downarrow\rangle+|b_\downarrow\rangle)+|c_\downarrow\rangle$. Via the $|c_\downarrow\rangle\ (=|\arrowspace\downarrow,\uparrow,\downarrow\rangle)$ - component of this state, it is possible to tunnel out with a majority ($\uparrow$ -) electron of the drain, yielding a transition to (g$_2)_{\downarrow\downarrow}$. Similarly, also starting from \geins{} in-tunneling of a $\downarrow$ - and subsequent out-tunneling of an $\uparrow$ - electron changes the $2n$ ground state to (g$_2)_{\downarrow\downarrow}$. Depending on the bias voltage, transport is either carried by $\uparrow$ - electron via the ground state channel (g$_2)_{\downarrow\downarrow}\leftrightarrow($t$_1)_\downarrow$, or by $\downarrow$ - electrons via (g$_2)_{\downarrow\downarrow}\leftrightarrow|\arrowspace\downarrow,\downarrow,\downarrow\rangle$, where $|\arrowspace\downarrow,\downarrow,\downarrow\rangle$ forms a blocking state unless a back-transition to the $2n$ excited state $|\arrowspace\downarrow,\downarrow\downarrow,\downarrow\rangle$ is energetically allowed.\\

\subsection{Tunneling magneto-resistance\label{sec:pol}}

\begin{figure}\begin{center}
\includegraphics[width=10.8cm]{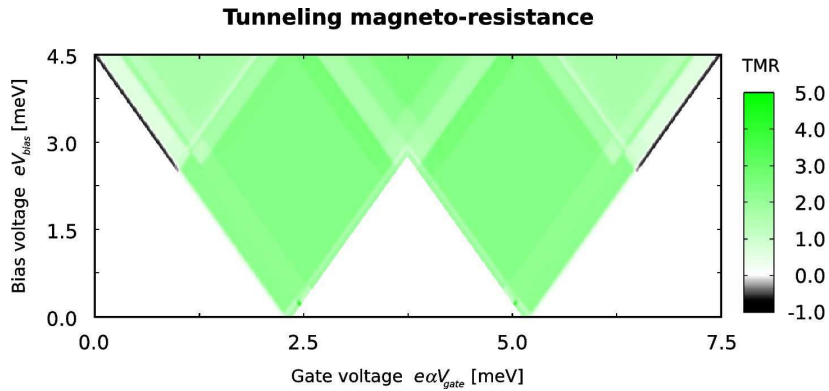}\vskip 0.2cm
\caption{tunneling magneto-resistance $I_{PA}/I_{AP}-1$, where $I_{PA/AP}$ denotes the current for parallel/anti-parallel polarized leads. We observe a negative value, i.e. $I_{PA}<I_{AP}$, along the edges of the $2n$/$2n+2$ Coulomb diamonds as soon as the channel \teins$\to$\eeins{} has opened. The reason is the decreased population of the \geins{} state in the anti-parallel case [\Fig{ACN-AP}b (right)] as compared to the unpolarized or parallel case [\Fig{ACN-80}a (right)]. \label{TMR-ACN}}
\end{center}
\end{figure}

In \Fig{TMR-ACN} we have plotted the tunneling magneto-resistance (TMR), \[\mathrm{TMR}:=\frac{I_{PA}-I_{AP}}{I_{AP}},\] which is a measure for the ratio of the current in the parallel configuration, $I_{PA}$ to the current in the anti-parallel configuration, $I_{PA}$. Along the edge of the $2n$ Coulomb diamond, the TMR acquires a negative value, i.e. $I_{AP}$ exceeds $I_{PA}$. This is unusual: for lowest order calculations without Zeeman splitting between the spin species typically strictly positive TMR is observed~\cite{Weymann05,Koller07,Hornberger08}. For the ACN, however, the effect originates from a reduced feeding of the \geins{} trapping state. This statement can be confirmed by comparing its occupation probability for the parallel, \Fig{ACN-80}a (right), and anti-parallel polarized case, \Fig{ACN-AP}b (right), in the concerned region.

Namely, \geins{} lies slightly lower in energy than \gdrei{} --\,for the values we chose, the energy difference amounts, according to \Tab{states-2n} and \Eq{xi}, to about $6\, k_BT$. Hence it can serve as a \emph{perfect} trapping state within a narrow region along the edge of the $2n$ Coulomb diamond: here, the bias is high enough to allow the ground state transition \gdrei{}$\to$\teins, but not \geins{}$\to$\teins{}. Though the latter channel is weak in any case, it nevertheless provides a nonzero escape rate from \geins. That is why in \Fig{ACN-80}a (right), in the region where the NDC mechanism \Fig{scheme} (II) can populate \geins{}, the occupation probability approaches $1$ only straight along the edge of the Coulomb diamond, and a value of $0.6-0.8$ further apart from it. In contrast, we observe no comparable increase of the \geins{} population in \Fig{ACN-AP}b (right), because for the anti-parallel configuration, as explained above, the transition channels involved in the NDC mechanisms are strongly disfavored compared to (g$_2)_{\downarrow\downarrow}\leftrightarrow($t$_1)_\downarrow$. For this reason, no trapping occurs and $I_{AP}$ can exceed $I_{PA}$, leading to the negative TMR.

\section{Magnetic field sweep\label{sec:mag}}
Finally we study, both for non-magnetic and collinearly polarized contacts (see Fig. \ref{graphene}), the transport behavior under the influence of an external magnetic field.
In its presence, formerly degenerate states with different spin projections $S_z$ components become Zeeman split.
This means, at a fixed gate voltage one half of the transitions occur at a higher, one half at a lower bias compared to the situation without magnetic field. In detail,
simple thoughts can confirm that the forward transitions involving $\uparrow$ - electrons, i.e. $2n\stackrel{\uparrow}{\to}2n+1$, as well as all backward transitions mediated by $\downarrow$ - electrons, i.e. $2n+1\stackrel{\downarrow}{\to}2n$,
are \emph{lowered} in bias, while  processes $2n\stackrel{\downarrow}{\to}2n+1$ and $2n+1\stackrel{\uparrow}{\to}2n$ are \emph{raised}.\\
\begin{figure}
\begin{center}\includegraphics[width=14cm]{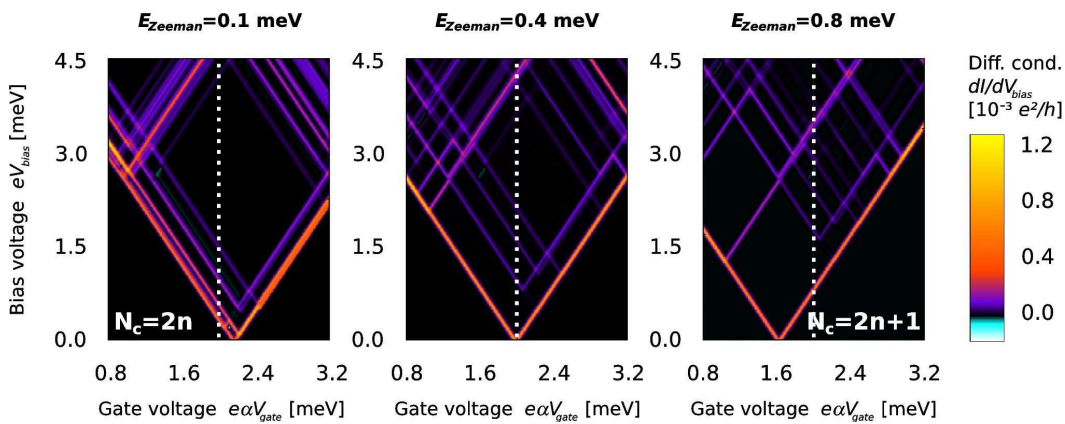}\end{center}
\caption{Differential conductance in the region between the $2n$ and $2n+1$ Coulomb blockade diamonds for an ACN quantum-dot at three different values of an external magnetic field. The contacts are assumed non-magnetic, i.e. unpolarized. The dashed white lines mark the gate voltage at which the plots \Fig{ACN-mag2} are taken.
\label{ACN-mag1}}
\end{figure}

\begin{figure}
\begin{center}\includegraphics[width=15.2cm]{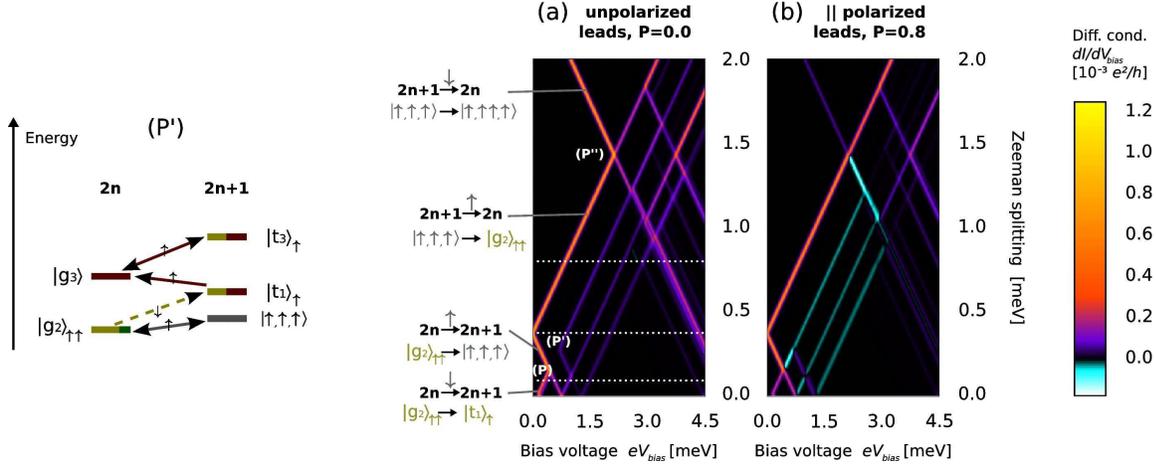}\end{center}
\caption{Differential conductance vs. bias voltage and Zeeman splitting at a fixed gate voltage of approx. $2.0\,$meV, for an ACN transport setup with (a) unpolarized and (b) parallel polarized contacts. The sketch (P') on the left explains why the channel (g$_2)_{\uparrow\uparrow}\to($t$_1)_{\uparrow}$ (dashed arrow) must be open in order to see transitions \gdrei$\leftrightarrow($t$_3)_\uparrow$.
\label{ACN-mag2}}
\end{figure}

For three distinct values of the Zeeman splitting, \Fig{ACN-mag1} shows stability diagram zooming on the region between the $2n$ and $2n+1$ Coulomb blockade diamonds. Those plots complement \Fig{ACN-mag2}a, where we show the differential conductance versus bias and Zeeman splitting, at a fixed gate voltage $e\alpha V_\gate\approx2.0\,$meV (marked in \Fig{ACN-mag1} with the dashed line). In turn, the three values of the Zeeman splitting considered in \Fig{ACN-mag1} are marked in \Fig{ACN-mag2}a by dashed white lines.

At first we focus on \Fig{ACN-mag2}a. In \Fig{ACN-mag2}b the case of polarized leads is considered.\\A small Zeeman splitting will actually select (g$_2)_{\uparrow\uparrow}$ as $2n$ ground state. The $2n\to2n+1$ ground-state-to-ground-state transition is then (g$_2)_{\uparrow\uparrow}{\to}(\mathrm{t}_1)_\uparrow$, as indicated left of the figure. It is the first line of a triple marking transitions to the lowest lying $2n+1$ states. Upon introducing a Zeeman energy, the spin-degeneracies of those are lifted, but only the two excited lines split in ``V''-like manner, while the ground state-to-ground state transition (g$_2)_{\uparrow\uparrow}{\to}(\mathrm{t}_1)_\uparrow$ has only one rightwards slanted branch (i.e. raises in energy with increasing field). The reason is that (g$_2)_{\uparrow\uparrow}$ is connected to the $|c_\uparrow\rangle\ (=|\arrowspace\uparrow,\downarrow,\uparrow\rangle)$ - component of the energetically favored state (t$_1)_\uparrow=\frac{1}{\sqrt{2}}(|a_\uparrow\rangle+|b_\uparrow\rangle)+|c_\uparrow\rangle$ by in-tunneling of $\downarrow$ - electrons. There is no possibility for a transition with $\uparrow$ - electrons, hence a left branch does not exist.\\
At a Zeeman splitting of $\sqrt{2}t/2$, the process (g$_2)_{\uparrow\uparrow}\stackrel{\uparrow}{\to}|\arrowspace\uparrow,\uparrow,\uparrow\rangle$ becomes the ground state-to-ground state transition. The crossover is marked with (P) in \Fig{ACN-mag2}a. Due to the \emph{in-tunneling} of $\uparrow$ - electrons, this resonance is continuously lowered in bias upon increasing the Zeeman energy further.\\
At a Zeeman splitting of about $0.4\,$meV, we are exactly at resonance. As seen in the middle plot of \Fig{ACN-mag1}, a line triple has clearly separated from the ground state transition line. Upon comparison with \Fig{ACN-mag2}a it is immediately understood that it belongs to $\downarrow$ - electron transitions to the lowest lying $2n+1$ states. Concerning the corresponding $\uparrow$ - electron transitions, something interesting happens: In the point (P'), the left branch of the second ``V''-shaped pattern, which belongs to the $2n+1$ state $(\mathrm{t}_3)_\uparrow$, ends. The reason is that $(\mathrm{t}_3)_\uparrow$ consists of the same components as $(\mathrm{t}_1)_\uparrow$. By in-tunneling of $\uparrow$ - electrons, it can thus not be connected to (g$_2)_{\uparrow\uparrow}$, but rather to \gdrei$\ [=\frac{1}{\sqrt{2}}(|a\rangle+|b\rangle)]$, see \Fig{ACN-mag2}, sketch (P'). This $2n$ state is, compared to (g$_2)_{\uparrow\uparrow}$, lifted by the Zeeman energy and can only be populated by back-transitions from (t$_1)_{\uparrow}$. The state (t$_1)_{\uparrow}$, however, is not available below the transition (g$_2)_{\uparrow\uparrow}\to($t$_1)_{\uparrow}$ [\Fig{ACN-mag2}, sketch (P'), dashed arrow]. This explains why the point (P') is positioned at the crossing with the resonance line marking this transition.\\
Going on to a value of $0.8\,$meV of the Zeeman splitting, where the rightmost plot in \Fig{ACN-mag1} is taken, we reside at low bias voltages within the $2n+1$ Coulomb blockade diamond; the ground state transition is now the \emph{out-tunneling} process $|\arrowspace\uparrow,\uparrow,\uparrow\rangle\stackrel{\uparrow}{\to}(\mathrm{g}_2)_{\uparrow\uparrow}$, thus raising in bias for an increasing magnetic field. The behavior reverts again in the point (P''), where the Zeeman splitting has lowered the excited $2n$ state $|\arrowspace\uparrow,\uparrow\uparrow,\uparrow\rangle$ enough to change the ground state transition to $|\arrowspace\uparrow,\uparrow,\uparrow\rangle{\to}|\arrowspace\uparrow,\uparrow\uparrow,\uparrow\rangle$, which involves out-tunneling of $\downarrow$ - electrons.\\
Finally, the \Fig{ACN-mag2}b shows the data obtained if the calculation yielding \Fig{ACN-mag2}a is performed for ferromagnetic leads, polarized in parallel to the applied field.
The only thing changing is the intensity of the lines. In particular, several of them are transformed into negative differential conductance lines. Such an effect is expected for any type of single electron transistor with parallel polarized contacts in magnetic field: upon opening a channel to a state from which the system can only escape via a weak (in our case $\downarrow$ - mediated) transition, NDC occurs as such slow processes hinder the current flow. The only exception are the ground state-to-ground state transitions, i.e. the edges of the Coulomb diamonds, where current starts to flow: to those, of course always positive differential conductance (PDC) lines belong. An obvious example is the $|\arrowspace\uparrow,\uparrow,\uparrow\rangle\stackrel{-\downarrow}{\to}|\arrowspace\uparrow,\uparrow\uparrow,\uparrow\rangle$ transition, which turns from NDC to PDC beyond the point (P'').

\section{Conclusion}
We have studied the transport characteristics of fully interacting graphene armchair nanoribbons (ACNs) attached to ferromagnetic contacts. Short-ranged Coulomb interactions play an essential role in such systems, leading to an entanglement between bulk states and the ones localized at the zig-zag ends of the ribbons, thereby lifting degeneracies between various states. Importantly, the entanglement breaks the otherwise strict conservation of the bulk spin-$S_z$ component, which leaves strong fingerprints in transport.\\
The stability diagrams predicted for ACNs possess a two-electron periodicity and show already for a completely symmetric, unpolarized setup in zero magnetic field unique features like a characteristic transition line triple and pronounced negative differential conductance \cite{Koller09}.\\These effects, originating from the interaction-induced lifting and formation of states symmetric or anti-symmetric under exchange of the ribbon ends, are preserved for a parallel contact polarization.
For an anti-parallel contact polarization, absence of various transition lines is observed due to spin-blockade effects and also the NDC features have vanished. The reason is that transition channels feeding the trapping state are disfavored, which leads even to a negative tunneling magneto-resistance.\\
We have further investigated the transport behavior in magnetic field, for unpolarized as well as for in parallel polarized contacts. A change of the odd filling ground state with $S_z=\hbar/2$ to one with $S_z=3\hbar/2$ is observed at a Zeeman splitting of $\sqrt{2}t/2$, such that the value of the end-bulk exchange coupling $t$ can directly be read off. Upon imposing a parallel contact magnetization, at several transition lines the differential conductance changes from the positive to the negative regime, because all $\downarrow$ - mediated transport channels become weak.\\

All in all, we found that short-ranged Coulomb interactions yield a strong influence of localized end states on the properties of ACNs. In particular, exchange makes the isolated bulk and end spin-$S_z$ components a bad quantum number: only the sum of both is a conserved quantity. Due to this fact, ACNs might not be as ideal candidates for certain spintronic devices as previously regarded. On the other hand, the entanglement is a rich source of ACN specific features in transport. Recent achievements in fabrication of carbon nanostripes with defined geometries~\cite{Yang08,Jioa09} raise the hope of an experimental confirmation of our predictions within the near future.

\ackn
We acknowledge the support of the DFG under the program SFB 689.
\appendix

\section{Diagonalization of $\hH_{0}+\hV_{\rho\rho}$\label{bosonization}}

We start by rewriting $\hH_{0}+\hV_{\rho\rho}$ in terms of collective
bosonic excitations. Concretely, we Fourier-expand the 1D electron
densities $\hrho_{r\sigma}(y)=\hpsi_{r\sigma}^{\dagger}(y)\hpsi_{r\sigma}(y),$
\begin{equation}
\hrho_{r\sigma}(y)=\frac{1}{2L_y}\sum_{q}e^{i rqy}\hrho_{\sigma q},\label{eq:rho_Fourier}\end{equation}
where the summation is over the wave numbers $q=\frac{\pi}{L_y}n_{q},\, n_{q}\in\mathbb{Z}.$
Then as shown e.g. in~\cite{Delft98} the operators \begin{equation}
\hb_{\sigma q}:=\frac{1}{\sqrt{n_{q}}}\hrho_{\sigma q}\,,\quad\hb^\dag_{\sigma q}:=\frac{1}{\sqrt{n_{q}}}\hrho_{\sigma -q},\quad q>0,\label{eq:Def-b}\end{equation}
fulfil the canonical bosonic commutation relations $[\hb_{\sigma q},\hb_{\sigma'q'}^{\dagger}]=\delta_{\sigma\sigma'}\delta_{qq'}.$\\As well known~\cite{Delft98} the bosonization of $\hH_{0}$ yields
\begin{equation}
\hH_{0}=\varepsilon_{0}\left(\sum_{\sigma}\sum_{q>0}n_{q}\hb_{\sigma q}^{\dagger}\hb_{\sigma q}+\frac{1}{2}\sum_{\sigma}\hat{\mathcal{N}}_{\sigma}^{2}\right),\label{eq:H_0_bosonized}\end{equation}
where $\hat{\mathcal{N}}_{\sigma}=\sum_{\kappa_y}\hc_{{\sigma\kappa}_{y}}^{\dagger}\hc_{{\sigma\kappa}_{y}}$ counts the number of electrons with
spin $\sigma$. The first term in Eq. (\ref{eq:H_0_bosonized}) accounts
for collective particle-hole excitations, whereas the second term is
due to Pauli's principle and describes the energy cost for filling
up the spin degenerate single-electron states. Terms proportional
to the total number of electrons have been omitted since they merely
lead to a shift of the chemical potential in transport experiments.\\Bosonization of $\hV^{{\bb}}_{\rho\rho}$
can be achieved by rewriting 
the involved potentials in terms of electron densities and inserting the Fourier expansion Eq. (\ref{eq:rho_Fourier}), thereby making use of the definition Eq. (\ref{eq:Def-b}). We obtain
\begin{eqnarray}
\hV^{{\bb}}_{\rho\rho}\nonumber & =&\hV^{{\bb}}_{f^+f^+}+\hV^{{\bb}}_{f^+f^-}+\hV^{{\bb}}_{f^- f^{+}}+\hV^{{\bb}}_{f^- f^{-}}+\hV^{{\bb}}_{b f^{+}}\\&=&\frac{1}{4}\sum_{\sigma\sigma'}\sum_{q}n_{q}W_{q}\left(\hb_{\sigma q}+\hb_{\sigma q}^{\dagger}\right)\left(\hb_{\sigma'q'}+\hb_{\sigma'q'}^{\dagger}\right) \nonumber\\
&& -  \frac{u}{4}\sum_{\sigma}\sum_{q>0}n_{q}\left(\hb_{\sigma q}\hb_{\sigma q}+\hb_{\sigma q}^{\dagger}\hb_{\sigma q}^{\dagger}\right) +\frac{1}{2}W_{0}\hat{\mathcal{N}}_{c}^{2} -  \frac{u}{4}\sum_{\sigma}\hat{\mathcal{N}}_{\sigma}^{2} ,\label{eq:V_rr_bos}\end{eqnarray}
with $W_{q}$ as given in \Eq{Wq}.\\
The last line of Eq. (\ref{eq:V_rr_bos}) describes the contribution of
$\hV^{{\bb}}_{\rho\rho}$ to the system energy depending on the number of electrons
in the two spin-bands. Here, $E_{c}=W_{0}$ is the ACN
charging energy; $\hat{\mathcal{N}}_{c}=\hat{\mathcal{N}}_{\uparrow}+\hat{\mathcal{N}}_{\downarrow}$
counts the total number of electrons. Spin alignment of electrons
 is favored by the term proportional
to $u$, because it counteracts the energy cost for the shell
filling in Eq.
(\ref{eq:H_0_bosonized}).\\
Finally, the bosonized form of the end-bulk contribution \Eq{V_e-b_rr} to $\hV_{\rho\rho}$ is
\begin{equation}\hV^{{\eb}}_{\rho\rho}=\frac{1}{4}\int \rmd y\ t_\rho(y)\,\sum_{\sigma}\sum_{q>0}\sqrt{n_q}\left(\hb_{\sigma q}+\hb^\dag_{\sigma q}\right)\cos(qy),\label{eq:V_eb_bos}\end{equation}
which is linear in the bosonic operators, while those appear quadratically in \Eqs{eq:H_0_bosonized}{eq:V_rr_bos}.\\In fact, any term of the form \Eq{eq:V_eb_bos} takes can be absorbed in the quadratic part of the Hamiltonian without any relevant impact on the spectrum,
and we remain with $\hH_{0}+\hV^{{\bb}}_{\rho\rho}$, which can be diagonalized in a standard way~\cite{Delft98} by a Bogoliubov transformation~\cite{Avery76}:
One introduces new bosonic operators $\ha_{j q}$ and $\ha_{j q}^{\dagger}$ which relate to the old bosonic operators, $\hb_{\sigma q}$ via
\begin{eqnarray}
&\hb_{\sigma q}=\sum_{j}\Lambda_\sigma^j\left(B_{jq}\ha_{jq}+D_{jq}\ha_{jq}^{\dagger}\right)\quad q>0\label{eq:b_a},\\
&\Lambda_{\sigma}^{c}=\frac{1}{\sqrt{2}}\,,\quad\Lambda_{\sigma}^{s}=\frac{\sgn(\sigma)}{\sqrt{2}}.
\label{eq:Lambdajd_rs}\end{eqnarray}
The transformation coefficients $B_{jq}$ and $D_{jq}$ can be expressed in terms of $X_{jq}$
and $A_{jq}$, which were introduced in \Sct{sec:diagdiag}:\[
B_{jq}=\frac{\varepsilon_{jq}+X_{jq}}{\sqrt{(\varepsilon_{jq}+X_{jq})^{2}-A_{q}^{2}}},\quad D_{jq}=-\frac{A_{jq}}{\sqrt{(\varepsilon_{jq}+X_{jq})^{2}-A_{q}^{2}}}.\]
With our values, \Eq{eq:X_xq}, for $X_{jq}$ and $A_{jq}$ we find approximately
\[
\varepsilon_{cq}=\varepsilon_{0}n_{q}\sqrt{1+{W_q}/{\varepsilon_{0}}},\quad
\varepsilon_{sq}=\varepsilon_{0}n_{q}\sqrt{1-\left({u}/{2\varepsilon_{0}}\right)^{2}}\approx\varepsilon_{0}n_{q},\]
and for the transformation coefficients to the spin mode
\begin{equation}
B_{sq}=1,\qquad D_{sq}=0.\label{eq:BsqDsq}\end{equation}
The transformation coefficients for the charge modes depend, as for SWCNTs~\cite{Mayrhofer08,Mayrhofer07}, on the ratio $g_{q}={\varepsilon_{0q}}/{\varepsilon_{cq}}$:
\begin{equation}
B_{cq}=\frac{1}{2}\left(\sqrt{g_q}+{1}/{\sqrt{g_{q}}}\right),\qquad D_{cq}=\frac{1}{2}\left(\sqrt{g_q}-{1}/{\sqrt{g_q}}\right).\label{eq:BcqDcq}\end{equation}
Exploiting these relations yields then the diagonalized Hamiltonian \Eq{eq:H0Vrr_diag}.

\section{The matrix elements of the non-diagonal bulk-bulk interaction\label{app:matrixelements-bulk}}
The evaluation of the non-diagonal bulk-bulk terms follows closely the procedure applied for SWCNTs~\cite{Mayrhofer08}.
In order to calculate the matrix elements of the non-diagonal bulk-bulk interaction
we have to derive an expression for the matrix element\begin{eqnarray*}
&\left\langle \vec{N}\vec{\sigma}^{{ \edge}}\vec{m}\left|\hV^{{\bb}}_{n\rho\rho}\right|\vec{N}'\vec{\sigma}^{{ \edge}}{}'\vec{m}'\right\rangle=:M_{[r][\sigma]_{f^-}}(\vec{N},\vec{\sigma}^{{ \edge}},\vec{m},\vec{N}',\vec{\sigma}^{{ \edge}}{}',\vec{m}',y)\\&:=\delta_{\vec{N},\vec{N}'}\delta_{\vec{\sigma}^{{ \edge}},\vec{\sigma}^{{ \edge}}{}'}\left\langle \vec{N}\vec{\sigma}^{{ \edge}}\vec{m}\left|\hpsi_{r_{1}\sigma}^{\dagger}(y)\hpsi_{r_{2}-\sigma}^{\dagger}(y)\hpsi_{r_{3}-\sigma}(y)\hpsi_{r_{4}\sigma}(y)\right|\vec{N}\vec{\sigma}^{{ \edge}}\vec{m}'\right\rangle\\&:=\delta_{\vec{N},\vec{N}'}\,\delta_{\vec{\sigma}^{{ \edge}},\vec{\sigma}^{{ \edge}}{}'}\,M_{[r][\sigma]_{f^-}}(\vec{N},y)M_{[r][\sigma]_{f^-}}(\vec{m},\vec{m}',y),\end{eqnarray*}which we have factorized in the last step into a fermionic and a bosonic part.
We express the operators $\hpsi_{r\sigma}(y)$ in
terms of the bosonic operators $\hb_{\sigma q}$ and $\hb_{\sigma q}^{\dagger},\, q>0$,
using the bosonization identity~\cite{Delft98}, \begin{equation}
\hpsi_{r\sigma F}(y)=\heta_{\sigma}\hK_{r\sigma}(y)e^{i\hphi_{r\sigma}^{\dagger}(y)}e^{i\hphi_{r\sigma}(y)}.\label{eq:bosident}\end{equation}
The operator $\heta_{\sigma}$ is the so called Klein factor, which
annihilates an electron in the $\sigma$-branch and thereby takes
care of the right sign as required from the fermionic anti-commutation
relations; in detail,
\begin{equation}
\heta_{\sigma}\left|\vec{N},\vec{m}\right\rangle =(-1)^{\delta_{\sigma,\downarrow}N_{\uparrow}}\left|\vec{N}-\vec{e}_{\sigma},\vec{m}\right\rangle .\label{eq:Def_Klein_factors}\end{equation}
$\hK_{r\sigma}(y)$ yields a phase factor depending on the number
of electrons of spin $\sigma$,\begin{equation}
\hK_{r\sigma}(y)=\frac{1}{\sqrt{2L_y}}e^{i\frac{\pi}{L_y}\mathrm{sgn}(r)(\hat{\mathcal{N}}_{\sigma}+\frac{1}{2})y}.\label{eq:K_rsF}\end{equation}
Finally, we have the boson fields $i\hphi_{r\sigma}(y)$, \begin{equation}
i\hphi_{r\sigma}(y)=\sum_{q>0}\frac{1}{\sqrt{n_{q}}}e^{i\mathrm{sgn}(r)qy}\hb_{\sigma q}.\label{eq:phifield_b}\end{equation}
The fermionic part is then given by
\begin{equation}
M_{[l]}(\vec{N},y)=
\left\langle \vec{N}\right|\hK_{l_{1}}^{\dagger}(y)\heta_{l_{1}}^{\dagger}\hK_{l_{2}}^{\dagger}(y)\heta_{l_{2}}^{\dagger}\hK_{l_{3}}(y)\heta_{l_{3}}\hK_{l_{4}}(y)\heta_{l_{4}}\left|\vec{N}'\right\rangle \end{equation}
and the bosonic part reads\begin{eqnarray}
&M_{[l]}(\vec{m},\vec{m}',y)=\left\langle \vec{m}\right|e^{-i \hphi_{l_{1}}^{\dagger}(y)}e^{-i \hphi_{l_{1}}(y)}e^{-i \hphi_{l_{2}}^{\dagger}(y)}e^{-i \hphi_{l_{2}}(y)}\nonumber\times\\&e^{i\hphi_{l_{3}}^{\dagger}(y)}e^{i\hphi_{l_{3}}(y)}e^{i\hphi_{l_{4}}^{\dagger}(y)}e^{i\hphi_{l_{4}}(y)}\left|\vec{m}'\right\rangle .\label{eq:Mmm_def_mt}\end{eqnarray}\smallskip

In order to improve readability we have replaced the indices $r\sigma$
by a single index $l$.
Using the relation (\ref{eq:Def_Klein_factors}) for the Klein factors $\heta_\sigma$, together with the fact that $S_\sigma=f^-$, $[\sigma]_{f^-}=[\sigma,-\sigma,-\sigma,\sigma]$, and the definition Eq. (\ref{eq:K_rsF}) of the phase factor $\hK_{r\sigma}(y)$, it is straightforward to show that
\begin{equation}
M_{[r][\sigma]_{f^-}}(\vec{N},y)=\frac{1}{(2L_y)^{2}}Q_{\vec{N}[r]\sigma}(y),\label{eq:MNN_TQ}\end{equation}
where $Q_{\vec{N}[r]\sigma}(y)=\exp{}\left[i\frac{\pi}{L_y}\Bigl(N_\sigma\,\sgn(r_4-r_1)-N_{-\sigma}\,\sgn(r_3-r_2)+\frac{\sgn(r_4+r_3-r_2-r_1)}{2}\Bigr)y\right]$.
Hence, for $S_r=u$, $[r]_u=[r,r,-r,-r]$, we obtain
\[
Q_{\vec{N}[r]_u\sigma}(y)=\exp{\left[-i\frac{2\pi}{L_y}\sgn(r)\left(N_\sigma+N_{-\sigma}+1\right)y\right]},
\]
which is oscillating fast with $N_c=N_\uparrow+N_\downarrow$ and thus completely suppresses the $S_r=u$ contribution away from half-filling. The only remaining term in $V^{{\bb}}_{n\rho\rho}$ is consequently $S_rS_\sigma=bf^-$, for which we get with $[r]_b=[r,-r,r,-r]$
\[
Q_{\vec{N}[r]_b\sigma}(y)=\exp{\left[-i\frac{2\pi}{L_y}\sgn(r)\left(N_\sigma-N_{-\sigma}\right)y\right]}.
\]
We can now restrict our further analysis to the bosonic part $M_{[r]_b[\sigma]_{f^-}}(\vec{m},\vec{m}',y)$.
We are going to express the fields $i\hphi_{r\sigma}(y)$ in Eq.
(\ref{eq:Mmm_def_mt}) in terms of the bosonic operators $\ha_{j q}$,
$\ha_{j q}^{\dagger}$ and subsequent normal ordering, i.e., commuting
all annihilation operators $\ha_{j q}$ to the right side and
all creation operators $\ha_{j q}^{\dagger}$ to the left side.
In a first step we use the relation
$e^{i\phi_{l}(y)}e^{i\hphi_{l}^{\dagger}(y)}=e^{i\hphi_{l}^{\dagger}(y)}e^{i\hphi_{l}(y)}e^{[i\hphi_{l}(y),i\hphi_{l}^{\dagger}(y)]}$,
following from the Baker-Hausdorff formula,
to obtain from Eq. (\ref{eq:Mmm_def_mt})\begin{equation}
M_{[l]}(\vec{m},\vec{m}',y)=C_{[l]}(y)
\left\langle \vec{m}\left|e^{-i \tilde{\sum}_{n=1}^{4}\hphi_{l_{n}}^{\dagger}(y)}e^{-i \tilde{\sum}_{n=1}^{4}\hphi_{l_{n}}(y)}\right|\vec{m}'\right\rangle ,\label{eq:Mmminterm_1}\end{equation}
where $\tilde{\sum}_{l=1}^{4}\hphi_{l_{n}}$ denotes the sum $\hphi_{l_{1}}+\hphi_{l_{2}}-\hphi_{l_{3}}-\hphi_{l_{4}}$
and \[
C_{[l]}(y)=e^{[i\hphi_{l_{3}}(y),i\hphi_{l_{4}}^{\dagger}(y)]}e^{[-i\hphi_{l_{2}}(y),i\hphi_{l_{3}}^{\dagger}(y)+i\hphi_{l_{4}}^{\dagger}(y)]}
e^{[-i\hphi_{l_{1}}(y),-i\hphi_{l_{2}}^{\dagger}(y)+i\hphi_{l_{3}}^{\dagger}(y)+i\hphi_{l_{4}}^{\dagger}(y)]}.\]
With the definition Eq. (\ref{eq:phifield_b}) of the boson fields, we can easily derive the anti-commutator relation \begin{equation}\left[i\hphi_{r\sigma}(y),i\hphi^\dag_{r'\sigma'}(y')\right]=-\delta_{\sigma,\sigma'}\sum_{q>0}{e^{i q(ry-r'y')}}/{n_q},\label{anticomm}\end{equation}
which allows us to simplify
$C_{[r]_b[\sigma]_{f^-}}(y)=\exp{[-i\hphi_{-r\sigma}(y),i\hphi_{r\sigma}^{\dagger}(y)]}\exp{[-i\hphi_{r\sigma}(y),i\hphi_{-r\sigma}^{\dagger}(y)]}=\left(1-\exp{[-2i r\pi y/L_y]}\right)^{-1}\left(1-\exp{[2i r\pi y/L_y]}\right)^{-1}=[4\sin^2\left({\pi}y/{L_y}\right)]^{-1}$.\\

Applying the Baker-Hausdorff formula once more, we obtain further for the second contribution to \Eq{eq:Mmminterm_1}\[
e^{-i \tilde{\sum}_{n=1}^{4}\hphi_{l_{n}}^{\dagger}(y)}e^{-i \tilde{\sum}_{n=1}^{4}\hphi_{l_{n}}(y)}=
e^{-i \tilde{\sum}_{n=1}^{4}\left(\hphi_{l_{n}}(y)+\hphi_{l_{n}}^{\dagger}(y)\right)}e^{\frac{1}{2}\left[i\tilde{\sum}_{n=1}^{4}\hphi_{l_{n}}^{\dagger}(y),i\tilde{\sum}_{n'=1}^{4}\hphi_{l_{n'}}(y)\right]}.\]
Using the definition Eq. (\ref{eq:phifield_b})
together with the transformation between the operators $\hat{b}_{\sigma q}$
and $\hat{a}_{j q}$, Eq. (\ref{eq:b_a}), we get\begin{equation}
i\hphi_{r\sigma}(y)+i\hphi_{r\sigma}^{\dagger}(y)=\sum_{j q>0}\left(\lambda_{r\sigma}^{j q}(y)\ha_{j q}-\lambda_{r\sigma}^{*j q}(y)\ha_{j q}^{\dagger}\right).\label{lmdasum}\end{equation}
In terms of $\Lambda_{\sigma}^{j}$, $B_{j q}$ and $D_{j q}$, which were introduced in \App{bosonization},
the coefficients $\lambda_{r\sigma}^{j q}(x)$ read\begin{equation}
\lambda_{r\sigma}^{j q}(y)=\frac{\Lambda_{\sigma}^{j}}{\sqrt{n_{q}}}\left(e^{i\mathrm{sgn}(r)qy}B_{j q}-e^{-i \mathrm{sgn}(r)qy}D_{j q}\right),\label{eq:lambda_einfach}\end{equation}
and plugging in the corresponding values, cf. Eqs. (\ref{eq:Lambdajd_rs}), (\ref{eq:BsqDsq}) and (\ref{eq:BcqDcq}), it is easy to calculate \begin{equation}
\tilde{\lambda}_{[l]}^{j q}(y):=-\tilde{\sum}_{n=1}^{4}\lambda_{l_{n}}^{j q}(y)\label{eq:lambda_tilde}\end{equation}
for $[l]=[r]_b[\sigma]_{f^-}$. We find that
\begin{eqnarray*}\tilde{\lambda}^{cq}_{[r]_b[\sigma]_{f^-}}(y)=0\,,\quad
\tilde{\lambda}^{sq}_{[r]_b[\sigma]_{f^-}}(y)=-2i\sqrt{\frac{2}{n_q}}\sgn(r\sigma)\Bigl(\underbrace{B_{sq}}_{\approx1}+\underbrace{D_{sq}}_{\approx0}\Bigr)\sin(qy).\end{eqnarray*}
Again using the Baker-Hausdorff formula yields for the exponential $e^{-i \tilde{\sum}_{n=1}^{4}\left(\hphi_{l_{n}}(y)+\hphi_{l_{n}}^{\dagger}(y)\right)}$\\$=\exp{\bigl(-\frac{1}{2}\sum_{q>0}\bigl|\tilde{\lambda}_{[r]_b[\sigma]_{f^-}}^{s q}(y)\bigr|^{2}\bigr)} e^{-\sum_{q>0}\tilde{\lambda}_{[r]_b[\sigma]_{f^-}}^{*s q}(y)\ha_{sq}^{\dagger}}e^{\sum_{q>0}\tilde{\lambda}_{[r]_b[\sigma]_{f^-}}^{s q}(y)\ha_{s q}}$,
such that in total \begin{eqnarray}\nonumber
\left\langle \vec{m}\left|e^{-i \tilde{\sum}_{n=1}^{4}\hphi_{r_{n}\sigma_n}^{\dagger}(y)}e^{-i \tilde{\sum}_{n=1}^{4}\hphi_{r_{n}\sigma_n}(y)}\right|\vec{m}'\right\rangle \\=\delta_{\vec{m}_c,\vec{m}'_c}
A_{[r]_b[\sigma]_{f^-}}(y)\prod_{q}F(\tilde{\lambda}_{[r]_b[\sigma]_{f^-}}^{s q}(y),m_{s q},m'_{sq}),\label{eq:Mmminterm_2}\end{eqnarray}
where we have introduced \[
A_{[r]_b[\sigma]_{f^-}}(y):=e^{\frac{1}{2}\left[i\tilde{\sum}_{n=1}^{4}\hphi_{r_{n}\sigma_{n}}^{\dagger}(y),i\tilde{\sum}_{n'=1}^{4}\hphi_{r_{n'}\sigma_{n'}}(y)\right]}e^{-\frac{1}{2}\sum_{q>0}\left|\tilde{\lambda}_{[r]_b[\sigma]_{f^-}}^{s q}(y)\right|^{2}}.\]
An explicit evaluation shows \[
A_{[r]_b[\sigma]_{f^-}}(y)=e^{\sum_{q>0}\frac{1}{n_q}\left(2-e^{-2i rqy}-e^{2i rqy}\right)} e^{-4\sum_{q>0}\frac{1}{n_q}\sin^2(qy)}=1.\]
The function $F(\lambda,m_{sq},m'_{sq})=\left\langle \vec{m}_s\left|e^{-\lambda^{*}\ha_{sq}^{\dagger}}e^{{\lambda}\ha_{s q}}\right|\vec{m}'_s\right\rangle $
is given by~\cite{Mayrhofer07,Mayrhofer08} \begin{eqnarray}
F(\lambda,m,m')=&
\left(\Theta(m'-m)\lambda^{m'-m}+\Theta(m-m')\left(-\lambda^{*}\right)^{m-m'}\right)\nonumber\\&
\times\sqrt{\frac{m_{min}!}{m_{max}!}}\sum_{l=0}^{m_{min}}\frac{\left(-\left|\lambda\right|^{2}\right)^{l}}{l!(l+m_{max}-m_{min})!}\frac{m_{max}!}{(m_{min}-l)!},\label{eq:Fvonlambda1}\end{eqnarray}
where $m_{min/max}=\min/\max(m,m')$. Combining \Eq{eq:Mmminterm_1}
and \Eq{eq:Mmminterm_2} we finally obtain\begin{equation}
M_{[r]_b[\sigma]_{f^-} }(\vec{m},\vec{m}',y)=\frac{\delta_{\vec{m}_c,\vec{m}'_c}}{4\sin^2({\pi}y/L_y)}\prod_{q}F(\tilde{\lambda}_{[r]_b[\sigma]_{f^-}}^{sq}(y),m_{s q},m'_{s q}).\\\label{appeq:boscontrib}\end{equation}
 Altogether, we get with \Eqs{eq:MNN_TQ}{appeq:boscontrib} to an expression for the matrix elements of $\hV^{{\bb}}_{n\rho\rho}$ away from half-filling,
\begin{eqnarray*}\nonumber\left\langle \vec{N}\vec{\sigma}^{{ \edge}}\vec{m}\left|\hV^{{\bb}}_{n\rho\rho}\right|\vec{N}'\vec{\sigma}^{{ \edge}}{}'\vec{m}'\right\rangle=\left\langle \vec{N}\vec{\sigma}^{{ \edge}}\vec{m}\left|\hV^{{\bb}}_{bf^-}\right|\vec{N}'\vec{\sigma}^{{ \edge}}{}'\vec{m}'\right\rangle \\=\frac{u}{2L_y} \delta_{\vec{N},\vec{N}'}\delta_{\vec{m}_c,\vec{m}'_c}\delta_{\vec{\sigma}^{{ \edge}},\vec{\sigma}^{{ \edge}}{}'}\sum_{r\sigma}\int \rmd y\frac{{Q}_{\vec{N}[r]_b\sigma}(y)}{4\sin^{2}\left(\pi y/{L_y}\right)}\prod_{q}F(\tilde{\lambda}_{[r]_b[\sigma]_{f^-}}^{s q}(y),m_{s q},m'_{s q}).\label{eq:ME_V_nrr-app}\end{eqnarray*}
For $\sum_{q}\left|m_{s q}-m'_{s q}\right|\le1$, the evaluation of this expression is problematic as the divergence arising from $1/[4\sin^{2}({\pi}y/{L_y})]$ remains uncompensated. Hence, the evaluation of the corresponding matrix elements needs
special care. The origin of this divergence lies in the
fact that, if no bosonic excitations are present, the $\vec{N}$ conserving
processes depend on the total number of electrons in the single branches
[compare to the fermionic contributions to $\hH_{0}+\hV_{\rho\rho}$ in
\Eq{eq:H0Vrr_diag}]. Since the bosonization approach requires
the assumption of an infinitely deep Fermi sea~\cite{Delft98}, this
leads, without the correct regularization, necessarily to divergences.
These findings are in complete analogy to the theory for SWCNTs~\cite{Mayrhofer08}.
In the following we exemplify the
proper calculation for $\left\langle \vec{N}\vec{m}\left|\hV^{{ \bb}}_{b\, f^{-}}\right|\vec{N}\vec{m}\right\rangle $.
\paragraph{Regularization of $M_{[r]_b[\sigma]_{f^-}}(\vec{N},\vec{\sigma}^{{ \edge}},\vec{m},\vec{N}',\vec{\sigma}^{{ \edge}}{}',\vec{m}',y)$ for $\vec{m}=\vec{m}'$\label{sec:regularization}}
Regularization for the matrix elements of the non-density-density bulk-bulk interaction is needed in case of $\sum_{jq}\left|m_{jq}-{m'}_{jq}\right|<2$,
since in that situation $M_{[r]_b[\sigma]_{f^-}}(\vec{N},\vec{\sigma}^{{ \edge}},\vec{m},\vec{N}',\vec{\sigma}^{{ \edge}}{}',\vec{m}',y)$
diverges due to the factor $1/4\sin^{2}(\frac{\pi}{L_y}y)$ in \Eq{appeq:boscontrib}.\\
Here we show the details of the proper regularization for $\vec{m}=\vec{m}'$. In this case we make the expansion \[
\prod_{q}F(\tilde{\lambda}^{sq}_{[r]_b[\sigma]_{f^-}}(y),\vec{m},\vec{m})=1+\mathcal{O}(\sin^{2}),\]
where $\mathcal{O}(\sin^2)$ contains only terms $\prod_{q}\left(\sin(qy)\right)^{t_{q}}$
with \mbox{$\sum_{q}t_{q}\ge2$} and which {}`cure' the $1/\sin^{2}({\pi}y/{L_y})$
divergence appearing in $M_{[r]_b[\sigma]_{f^-}}(\vec{N},\vec{m},\vec{m},y)$.
Therefore we are, compare to Eq. (\ref{eq:ME_V_nrr-app}), left with the regularization of
\[
\frac{{Q}_{\vec{N}[r]_b\sigma}(y)}{4\sin^{2}\left(\pi y/L_y\right)}=\frac{e^{-i \frac{2\pi}{L_y}r\left(N_\sigma-N_{-\sigma}\right)y}}{4\sin^{2}\left({\pi}y/L_y\right)}=\frac{e^{-2i \frac{\pi}{L_y}rN_{\sigma}y}}{1-e^{i r\frac{2\pi}{L_y}y}}\frac{e^{2i \frac{\pi}{L_y}rN_{-\sigma}y}}{1-e^{-i r\frac{2\pi}{L_y}y}},
\]
with the second equality obtained following~\cite{Mayrhofer08}. Using further
$\sum_{n=-\infty}^{N}e^{-i nx}=\frac{e^{-i Nx}}{1-e^{i x}},$
this leads to\begin{equation*}\frac{{Q}_{\vec{N}[r]_b\sigma}(y)}{4\sin^{2}\left(\pi y/L_y\right)}=
\sum_{n=-\infty}^{N_{\sigma}}e^{-i nr\frac{2\pi}{L_y}y}\sum_{n'=-\infty}^{N_{-\sigma}}e^{i n'r\frac{2\pi}{L_y}y}.\label{eq:sintosums}\end{equation*}
Integration over $y$ brings us to \[
\int\! \rmd y\,\frac{{Q}_{\vec{N}[r]_b\sigma}(y)}{4\sin^{2}\left(\pi y/L_y\right)}=\sum_{n=-\infty}^{N_{\sigma}}\sum_{n'=-\infty}^{N_{-\sigma}}\!\!L_y\,\delta_{n,n'}=L_y \min(N_{\sigma},N_{-\sigma}).\]
Summarizing, the regularized expression reads:
\begin{eqnarray*}\left\langle \vec{N}\vec{\sigma}^{{ \edge}}\vec{m}\left|\hV^{{\bb}}_{bf^-}\right|\vec{N}'\vec{\sigma}^{{ \edge}}\vec{m}\right\rangle= u\,\delta_{\vec{N},\vec{N}'}\delta_{\vec{\sigma}^{{ \edge}},\vec{\sigma}^{{ \edge}}{}'}
\left[ \sum_{\sigma} \min(N_{\sigma},N_{-\sigma})\right.\nonumber\\+\left.\frac{1}{2L_y}\sum_{r\sigma}\int \rmd y\,\frac{{Q}_{\vec{N}[r]_b\sigma}(y)}{4\sin^{2}\left({\pi}y/L_y\right)}\prod_{q}\left(F(\tilde{\lambda}_{[r]_b[\sigma]_{f^-}}^{s q}(y),m_{s q},m_{s q})-1\right)\right].\label{eq:ME_V_nrr_expl}\end{eqnarray*}

\section{\label{sec:CalcVeb}The matrix element of the non-diagonal end-bulk interaction}

Also for the non-density-density end-bulk-scattering, we omitted the calculations in the main part of the text and give the detailed evaluations here.
We have to evaluate matrix elements of Eq. (\ref{V_e-b_nrr}), of the form
$M^p_{r\sigma r'\sigma'}(\vec{N},\vec{\sigma}^{{ \edge}},\vec{m},\vec{N}',\vec{\sigma}^{{ \edge}}{}',\vec{m}'):=
\left\langle \vec{N}\vec{\sigma}^{{ \edge}}\vec{m}\left|\hpsi_{r\sigma}^{\dagger}(y_p)\hpsi_{r'\sigma'}(y_p)\sum\nolimits_{\kappa_x}\hd^{\dagger}_{\sigma'p\kappa_x}\hd_{\sigma p\kappa_x}\right|\vec{N}'\vec{\sigma}^{{ \edge}}{}'\vec{m}'\right\rangle$, 
with $y_{-}$=$0,\,y_{+}$=$L_y$ as employed previously. We can factorize the matrix elements into a fermionic, a bosonic, and an end part\[
M^{p}_{r\sigma r'\sigma'}(\vec{N},\vec{\sigma}^{{ \edge}},\vec{m},\vec{N'},\vec{\sigma}^{{ \edge}}{}',\vec{m}')
=M^p_{r\sigma r'\sigma'}(\vec{N},\vec{N}')M^p_{r\sigma r'\sigma'}(\vec{m},\vec{m}')M^p_{\sigma\sigma'}(\vec{\sigma}^{{ \edge}},\vec{\sigma}^{{ \edge}}{}').\]
\noindent The easiest to give is the end contribution:
\begin{equation}
M^p_{\sigma\sigma'}(\vec{\sigma}^{{ \edge}},\vec{\sigma}^{{ \edge}}{}')=\sum_{\kappa_x}\left\langle \vec{\sigma}^{{ \edge}}\left|d^{\dagger}_{\sigma'p\kappa_x}d_{\sigma p\kappa_x}\right|\vec{\sigma}^{{ \edge}}{}'\right\rangle=\delta_{\sigma^{\small \edge}_{\bar{p}},\sigma_{\bar{p}}^{\small \edge}{}'}\,\delta_{\sigma^{\small \edge}_p,\sigma'}\,\delta_{\sigma^{\small \edge}_p{}',\sigma}.\label{Meb-edge}\end{equation}
The end electron operators act on the configuration at the $p$-end only, trying to transform the spin from $\sigma$ to $\sigma'$, while the spin at $\bar{p}=-p$ must be untouched.

What remains is $\langle \vec{N}\vec{m}|\hpsi_{r\sigma}^{\dagger}(y_p)\hpsi_{r'\sigma'}(y_p)|\vec{N}'\vec{m}'\rangle$. Upon applying the bosonization identity Eq. (\ref{eq:bosident}) to rewrite $\hpsi_{r\sigma}(y_p)$, it becomes a product of the fermionic part $M^p_{r\sigma r'\sigma'}(\vec{N},\vec{N'})$, which is calculated below, and the bosonic part,
\begin{eqnarray*}M^p_{ll'}(\vec{m},\vec{m}')=\left< \vec{m} \left|e^{-i\hphi^\dagger_{l}(y_p)}e^{-i \hphi_{l}(y_p)}e^{i \hphi^\dagger_{l'}(y_p)}e^{i \hphi_{l'}(y_p)}\right|\vec{m}'\right>\\=C_{ll'}(y_p) A_{ll'}(y)\prod_{j,q>0}F(\tilde{\lambda}_{ll'}^{jq}(y),m_{jq},m'_{jq}),\end{eqnarray*}
where we abbreviated $r\sigma$ by $l$ and  $r'\sigma'$ by $l'$ and exploited \Eq{anticomm} to obtain
$C_{ll'}(y):=e^{\left[-i\hphi_{l}(y),i\hphi^\dagger_{l'}(y)\right]}=e^{\delta_{\sigma,\sigma'}\sum_{q>0}\frac{1}{n_q}e^{i (r-r')qy}}=e^{\delta_{\sigma,\sigma'}\sum_{q>0}\frac{1}{n_q}\left\{\cos(2\delta_{r,-r'}qy)+i\sgn(r)\sin(2\delta_{r,-r'}qy)\right\}}$. 
For the remaining exponentials we did as for the bulk-bulk scattering in \App{app:matrixelements-bulk}, and while $F(\lambda,m,m')$ is the function defined in \App{app:matrixelements-bulk}, \Eq{eq:Fvonlambda1},
$A_{ll'}(y):=e^{\frac{1}{2}\left[i\left(\hphi^\dagger_{l}(x)-\hphi^\dagger_{l'}(x)\right),i\left(\hphi_{l}(x)-\hphi_{l'}(x)\right)\right]}e^{-\frac{1}{2}\sum_{jq}\left|\tilde{\lambda}_{ll'}^{jq}\right|^2}=e^{\sum_{q>0}\frac{1}{n_q}(1-\delta_{\sigma,\sigma'}\cos{(2\delta_{r,r'}qy)})}e^{-\frac{1}{2}\sum_{jq}\left|\tilde{\lambda}_{ll'}^{jq}(y)\right|^2}$
such that
\begin{equation}
A_{ll'}(y)C_{ll'}(y)=e^{\sum_{q>0}\frac{1}{n_q}\left(1+i\delta_{r,-r'}\sgn(r)\sin(2qy)\right)}e^{-\frac{1}{2}\sum_{jq}\left|\tilde{\lambda}_{ll'}^{jq}\right|^2}. \label{AC}\end{equation}
Further, we set in \Eq{eq:lambda_einfach} to expand
$\tilde{\lambda}_{ll'}^{jq}(y):=-\lambda^{jq}_{l}(y)+\lambda^{jq}_{l'}(y)
=\frac{1}{\sqrt{n_q}}\left[B_{jq}\left(\Lambda^j_{\sigma'}e^{i \sgn(r')qy}-\Lambda^j_{\sigma}e^{i \sgn(r)qy}\right)-D_{jq}\left(\Lambda^j_{\sigma'}e^{-i \sgn(r')qy}-\Lambda^j_{\sigma}e^{-i \sgn(r)qy}\right)\right]$. 
For our matrix element $M^p_{ll'}(\vec{m},\vec{m}')$ we have only to consider the two cases $y=y_-=0$ and $y=y_+=L_y$,
where the expression simplifies under application of \Eqs{eq:Lambdajd_rs}{eq:BsqDsq} along with $\sin(0)=\sin(\pm qL_y)=0,\ \cos(0)=1,\ \cos(\pm qL_y)=-1$ to
\begin{equation*}\tilde{\lambda}_{ll'}^{jq}(y_p)=\delta_{j,s}\delta_{\sigma,-\sigma'}\sgn(p\sigma)\sqrt{{2}/{n_q}}.\end{equation*}
Using this result in \Eq{AC} yields
\begin{equation*}
A_{ll'}(y_p)C_{ll'}(y_p)=\left\{\begin{array}{lr}const.&\sigma'=\sigma,\\1&\sigma'=-\sigma.\end{array}\right.\end{equation*}
With this ingredients we arrive eventually at the crucial result
\begin{equation}M^p_{ll'}(\vec{m},\vec{m}')=\left\{\begin{array}{lr}const.&\sigma'=\sigma,\\\delta_{\vec{m}_c,\vec{m}'_c}\prod_{q>0}F(\sgn(p\sigma)\sqrt{{2}/{n_q}},m_{sq},m'_{sq})&\sigma'=-\sigma.\end{array}\right.\label{Meb_mm}\end{equation}

Finally, the fermionic part is given by
\begin{eqnarray}
&M^p_{r\sigma r'\sigma'}(\vec{N},\vec{N'})=\left\langle \vec{N}\right|\hK_{r\sigma}^{\dagger}(y_p)\heta_{\sigma}^{\dagger}\heta_{\sigma'}\hK_{r'\sigma'}(y_p)\left|\vec{N}'\right\rangle=\nonumber\\&\frac{1}{2L_y}\delta_{\vec{N},\vec{N}'+\vec{e}_{\sigma}-\vec{e}_{\sigma'}}(-1)^{(1-\delta_{\sigma,\sigma'})(N'_\uparrow-\delta_{\sigma,\downarrow})}e^{i \frac{\pi}{L_y}\left[r'\left(N'_{\sigma'}+\frac{1}{2}\right)-r\left(N'_{\sigma}+\frac{3}{2}\right)\right]y_p},\label{Meb-ferm}
\end{eqnarray}
where the Klein factors $\hat{\eta}_{\sigma}$, \Eq{eq:Def_Klein_factors}, and the phase factors $\hat{K}_{r\sigma}(y)$, \Eq{eq:K_rsF}, were straightforward to evaluate.
We split now the sum $\sum\nolimits_{\sigma\sigma'}$ contained in $\hV^{{\eb}}_{n\rho\rho}$, \Eq{V_e-b_nrr}, and therewith the further analysis, into two cases.\\
First, let $\sigma'=\sigma$ hold. Then,
\[
M^p_{r\sigma r'\sigma}(\vec{N},\vec{N'})=\left\{\begin{array}{ll}{\delta_{\vec{N},\vec{N'}}}/\left({2L_y}\right)&\ y_p=y_-=0\,,\\-rr'\,{\delta_{\vec{N},\vec{N'}}}/\left({2L_y}\right)&\ y_p=y_+=L_y\,.\end{array}\right.
\]
The total contribution related to this case is ($t\equiv t_{n\rho}$)
\[\frac{t}{4}\delta_{\vec{N},\vec{N}'}\sum_{\sigma}\sum_{rr'}\sum_{p}\left(\delta_{p,+}-\delta_{p,-}\right)M^p_{r\sigma r'\sigma}(\vec{m},\vec{m}')\delta_{\sigma^{\small \edge}_{\bar{p}},\sigma_{\bar{p}}^{\small \edge}{}'}\,\delta_{\sigma,\sigma^{\small \edge}_p}\,\delta_{\sigma,\sigma^{\small \edge}_p{}'}
.\]
According to \Eq{Meb_mm}, in fact $M^+_{r\sigma r'\sigma}(\vec{m},\vec{m}')=M^-_{r\sigma r'\sigma}(\vec{m},\vec{m}')\equiv const.$ and so this contribution identically vanishes.\\
We are left with the part of the sum where $\sigma'\neq\sigma$, for which we can read off the following fermionic contributions from \Eq{Meb-ferm}:
\begin{equation}
M^{p}_{r\sigma r'-\sigma}(\vec{N},\vec{N'})=\frac{\delta_{\vec{N},\vec{N'}+\vec{e}_\sigma-\vec{e}_{-\sigma}}}{2L_y}\sgn(\sigma)\left(\delta_{p,-}(-1)^{N'_\uparrow}-\delta_{p,+}rr'(-1)^{N'_\downarrow}\right),\label{Meb-ferm-final}\end{equation}
To obtain the first equality we used $(-1)^{N'_\uparrow-\delta_{\sigma,\downarrow}}=-(-1)^{N'_\uparrow}\sgn(\sigma)$, while for the second one we needed $\exp{\{i\pi[r'(N'_{-\sigma}+0.5)-r(N'_\sigma+1.5)]\}}=-\sgn(rr')(-1)^{N'_\sigma}(-1)^{N'_{-\sigma}}$.\\
Putting Eqs. (\ref{Meb-edge}), (\ref{Meb_mm}) and (\ref{Meb-ferm-final}) together, the final result for the non-diagonal end-bulk-scattering is:
\begin{eqnarray*}\nonumber&\left\langle \vec{N}\ \vec{\sigma}^{{ \edge}}=(\sigma_+,\sigma_-)\ \vec{m}\left|\hV^{{\eb}}_{n\rho\rho}\right|\vec{N}'\ \vec{\sigma}^{{ \edge}}{}'=(\sigma'_+,\sigma'_-)\ \vec{m}'\right\rangle
\\\nonumber&=t\sum_{p}\delta_{\vec{N},\vec{N'}+\vec{e}_{\sigma'_p}-\vec{e}_{\sigma_p}}\delta_{\vec{m}_c,\vec{m}'_c}\delta_{\sigma_p,-\sigma'_{p}}\delta_{\sigma_{\bar{p}},\sigma'_{\bar{p}}}\,\left[\delta_{p,-}(-1)^{N'_\uparrow}-\delta_{p,+}(-1)^{N'_\downarrow}\right]\\\ &\times\mathrm{sgn}(\sigma_p)
\prod_qF\left(p\,\mathrm{sgn}(\sigma'_p)\sqrt{2/n_q},m_{sq},m'_{sq}\right).\end{eqnarray*}
\section*{References}
\bibliography{paper}

\providecommand{\newblock}{}
\begin{thebibliography}{10}
\expandafter\ifx\csname url\endcsname\relax
  \def\url#1{{\tt #1}}\fi
\expandafter\ifx\csname urlprefix\endcsname\relax\def\urlprefix{URL }\fi
\providecommand{\eprint}[2][]{\url{#2}}

\bibitem{Novoselov04}
Novoselov K~S, Geim A~K, Morozov S~V, Jiang D, Zhang Y, Dubonos S~V, Grigorieva
  I~V and Firsov A~A 2004 {\em Science\/} {\bf 306} 666

\bibitem{Falkol07}
Fal'ko1 V 2007 {\em Nature Physics\/} {\bf 3} 151

\bibitem{Chen08}
Chen J~H, Jang C, Xiao S, Ishigami M and Fuhrer M~S 2008 {\em Nature
  Nanotechnology\/} {\bf 3} 206

\bibitem{BeenakkerCOL}
Beenakker C~W~J 2008 {\em Rev.\ Mod.\ Phys.\/} {\bf 80} 1337

\bibitem{Wallace47}
Wallace P~R 1947 {\em Phys.\ Rev.\/} {\bf 71} 622

\bibitem{Zhang05}
Zhang Y, Tan Y~W, Stormer H~L and Kim P 2005 {\em Nature\/} {\bf 438} 201

\bibitem{Zhou06}
Zhou S~Y, Gweon G~H, Graf J, Federov A~V, Spatura C~D, Diehl R~D, Kopelevich Y,
  Lee D~H, Louie S~G and Lanzara A 2006 {\em Nat. Phys.\/} {\bf 2} 595

\bibitem{Bostwick07}
Bostwick A, Ohta T, Seyller T, Horn H~K and Rotenberg E 2007 {\em Nat. Phys.\/}
  {\bf 3} 36

\bibitem{Han07}
Han J~E and Heary R~J 2007 {\em Physical Review Letters\/} {\bf 99} 236808

\bibitem{Stampfer08}
Stampfer C, G\"uttinger J, Hellm\"uller S, Molitor F, Ensslin K and Ihn T 2008
  {\em Nano\ Lett.\/} {\bf 8} 2378

\bibitem{Lin08}
Lin Y~M, Perebeinos V, Chen Z and Avouris P 2008 {\em Phys.\ Rev.\ B\/} {\bf
  78} 161409(R)

\bibitem{Schnez09}
Schnez S, Molitor F, Stampfer C, G\"uttinger J, Shorubalko I, Ihn T and Ensslin
  K 2009 {\em Appl.\ Phys.\ Lett.\/} {\bf 94} 012107

\bibitem{Molitor09}
Molitor F, Dr\"oscher S, G\"uttinger J, Jacobsen A, Stampfer C, Ihn T and
  Ensslin K 2009 {\em Appl.\ Phys.\ Lett.\/} {\bf 94} 222107

\bibitem{Moriyama09}
Moriyama S, Tsuya D, Watanabe E, Uji S, Shimizu M, Mori T, Yamaguchi T and
  Ishibashi K 2009 {\em Nano\ Lett.\/} {\bf 9} 2891

\bibitem{Sols07}
F~Sols F~G and Castro-Neto A~H 2007 {\em Phys.\ Rev.\ Lett.\/} {\bf 99} 166803

\bibitem{Zarea07}
Zarea M and Sandler N 2007 {\em Phys.\ Rev.\ Lett.\/} {\bf 99} 256804

\bibitem{Tapaszto08}
Tapaszto L, Dobrik G, Lambin P and Biro L 2008 {\em Nature Nanotechnology\/}
  {\bf 3} 397

\bibitem{Yang08}
Yang X, Dou X, Rouhanipour A, Zhi L, R\"ader H~J and M\"ullen K 2008 {\em
  JACS\/} {\bf 130} 4216

\bibitem{Jioa09}
Jiao L, Zhang L, Wang X, Diankov G and Dai H 2009 {\em Nature\/} {\bf 458} 877

\bibitem{Nakada96}
Nakada K, Fujita M, Dresselhaus G and Dresselhaus M~S 1996 {\em Phys.\ Rev.\
  B\/} {\bf 54} 17954

\bibitem{Akhmerov08}
Akhmerov A~R and Beenakker C~W~J 2008 {\em Phys.\ Rev.\ B\/} {\bf 77} 085423

\bibitem{Fujita96}
Fujita M, Wakabayashi K, Nakada K and Kusakabe K 1996 {\em J. Phys. Soc.
  Jpn.\/} {\bf 65} 1920

\bibitem{Klusek05}
Klusek Z, Kozlowski W, Waqar Z, Datta S, Burnell-Gray J~S, Makarenko I~V, Gall
  N~R, Rutkov E~V, Tontegode A~Y and Titko A 2005 {\em Applied Surface
  Science\/} {\bf 252} 1221

\bibitem{Kobayashi06}
Kobayashi Y, Fukui K, Enoki T and Kusakabe K 2006 {\em Phys.\ Rev.\ B\/} {\bf
  73} 125415

\bibitem{Niimi06}
Niimi Y, Matsui T, Kambara H, Tagami K, Tsukada M and Fukuyama H 2006 {\em
  Phys.\ Rev.\ B\/} {\bf 73} 085421

\bibitem{Son06}
Son Y~W, Cohen M~L and Louie S~G 2006 {\em Nature\/} {\bf 444} 347

\bibitem{Wimmer08}
Wimmer M, Adaglideli I, Berber S, Tom\'anek D and Richter K 2008 {\em Phys.\
  Rev.\ Lett.\/} {\bf 100} 177207

\bibitem{Wunsch08}
Wunsch B, Stauber T, Sols F and Guinea F 2008 {\em Phys.\ Rev.\ Lett.\/} {\bf
  101} 036803

\bibitem{Koller09}
Koller S, Mayrhofer L and Grifoni M 2009 {\em EPL\/} {\bf 88} 57001

\bibitem{Castro-Neto08}
Castro-Neto A~H, Guinea F, Peres N~M~R, Novoselov K~S and Geim A~K 2009 {\em
  Rev.\ Mod.\ Phys.\/} {\bf 81} 109

\bibitem{Brey06}
Brey L and Fertig H~A 2006 {\em Phys.\ Rev.\ B\/} {\bf 73} 235411

\bibitem{Barford05}
Barford W 2005 {\em Electronic and Optical Properties of Conjugated Polymers\/}
  (Oxford: Clarendon Press)

\bibitem{Fulde95}
Fulde P 1995 {\em Electron Correlations in Molecules and Solids\/} (Berlin:
  Springer)

\bibitem{Egger97}
Egger R and Gogolin A~O 1997 {\em Phys.\ Rev.\ Lett.\/} {\bf 79} 5082

\bibitem{Mayrhofer08}
Mayrhofer L and Grifoni M 2008 {\em Eur. Phys. J. B\/} {\bf 63} 43

\bibitem{DiEl}
A dielectric constant $\epsilon=1.4$ was assumed.

\bibitem{Mayrhofer06}
Mayrhofer L and Grifoni M 2006 {\em Phys.\ Rev.\ B\/} {\bf 74} 121403(R)

\bibitem{Koller07}
Koller S, Mayrhofer L and Grifoni M 2007 {\em New J. Phys.\/} {\bf 9} 348

\bibitem{Hornberger08}
Hornberger R~P, Koller S, Begemann G, Donarini A and Grifoni M 2008 {\em Phys.\
  Rev.\ B\/} {\bf 77} 245313

\bibitem{Mayrhofer07}
Mayrhofer L and Grifoni M 2007 {\em Europ. Phys. J. B\/} {\bf 56} 107

\bibitem{Weymann05}
Weymann I, K{\"o}nig J, Martinek J, Barnas J and Sch{\"o}n G 2005 {\em Phys.\
  Rev.\ B\/} {\bf 72} 115334

\bibitem{Delft98}
v~Delft J and Schoeller H 1998 {\em Annalen Phys.\/} {\bf 7} 225

\bibitem{Avery76}
Avery J 1976 {\em Creation and Annihilation Operators\/} (McGraw-Hill, New
  York)

\end{thebibliography}
\bibliographystyle{iopart-num}
\end{document}